\documentstyle[11pt]{article}
\begin{document}
\begin{titlepage}
\begin{flushright}
YITP-00-39\\
hep-th/0007112\\
July 2000
\end{flushright}
\begin{centering}
 
{\ }\vspace{1cm}
 
{\Large\bf Half-Integer Winding Number Solutions}\\

\vspace{10pt}

{\Large\bf to the Landau-Ginzburg-Higgs Equations}\\

\vspace{10pt}

{\Large\bf and Instability of the Abrikosov-Nielsen-Olesen Vortex}\\

\vspace{2.0cm}

Jan Govaerts\footnote{On leave from the Institute of Nuclear
Physics, Catholic University of Louvain, Louvain-la-Neuve, Belgium\\
\indent{\ }\hspace{5pt}E-mail: {\tt govaerts@fynu.ucl.ac.be}}\\

\vspace{0.6cm}

{\em C.N. Yang Institute for Theoretical Physics}\\
{\em State University of New York at Stony Brook}\\
{\em Stony Brook NY 11794-3840, USA}\\

\vspace{1.5cm}
\begin{abstract}

\noindent New solutions to the abelian U(1) Higgs model, corresponding
to vortices of integer and half-integer winding number bound onto
the edges of domain walls and possibly surrounded by annular 
current flows, are described, based on a fine-grained analysis
of the topology of such configurations in spacetime. The existence
of these states, which saturate BPS bounds in specific limits and are
quite reminiscent of D-branes and membranes in general, could have
interesting and some important consequences in a wide range of physical
contexts. For instance, they raise the possibility that for some regimes
of couplings the usual vortex of unit winding number would split into
two vortices each of one-half winding number bound by a domain wall.
A similar approach may also be relevant to other known topological states
of field theory.

\end{abstract}

\vspace{10pt}

\end{centering} 

\vspace{5pt}


\vspace{25pt}

\end{titlepage}

\setcounter{footnote}{0}

\section{\bf Introduction.}

The pivotal role played by what is now called the abelian U(1) Higgs model 
in the progress and the stimulus for new ideas in many areas of physics over 
the second half of the twentieth century need not be emphasized\cite{Weinberg}.
Ranging from phase transitions in condensed matter systems displaying 
quantum coherence phenomena such as superconductivity and superfluidity,
through elementary particle physics with its issues of spontaneous and 
dynamical symmetry breaking, the origin of mass, colour confinement 
and the dual Meissner effect, to cosmology and the evolution of the 
universe with the formation of textures of different dimensionalities 
all possibly within some inflationary scenario, and finally even M-theory 
to the extent that magnetic vortex solutions provided the first example 
of so-called BPS states which are so crucial to the duality web of 
superstring theories, the abelian Higgs model has proved to be sort of 
a precious treasure chest for tools in our quest of the secrets of the 
entire physical universe, cast away that we are on our little 
``speckle of dust". With Dirac's monopole, magnetic vortices were also the 
first known states bearing witness to the wide-ranging role played by the 
topology of field configurations in classical and quantum field theories 
alike, thereby opening up yet another whole beautiful toolbox of 
mathematical physics. 

In this latter respect, it seems to be {\sl de facto\/}
a well accepted working assumption---albeit not established in
any strict mathematical sense---that all topologically non trivial 
solutions to the Landau-Ginzburg-Higgs (LGH) equations are magnetic 
vortices of integer winding number in the order parameter represented 
by a spacetime dependent complex scalar field,
\begin{equation}
\psi(x)=f(x)e^{i\theta(x)}.
\label{eq:EQ1}
\end{equation}
Indeed, in a space of infinite extent, finite energy configurations
must necessarily be such that $|\psi(x)|$ approaches its non vanishing
constant vacuum expectation value at infinity, while single-valuedness 
of $\psi(x)$ itself then allows only for a phase dependency of integer 
winding number $L$ at infinity, $e^{-iL\phi}$ ($\phi$ being the angular 
direction in a plane locally transverse to the vortex). Continuity of 
$\psi(x)$ throughout space then implies (when $L\ne 0$) that $\psi(x)$ 
itself must vanish along at least one curve of dimension one at a finite 
distance which determines the location in three dimensional space
of at least one vortex. Since such configurations carry a non vanishing 
magnetic flux measured at infinity 
which is directly proportional to the winding number $L$,
they indeed correspond to magnetic vortices. In particular, the $L=\pm 1$ 
solution is the celebrated Abrikosov (anti)vortex of superconductors
described by the Landau-Ginzburg equations\cite{Abrikosov}, or equivalently the 
Nielsen-Olesen vortex of the abelian U(1) Higgs model whose low energy
effective relativistic dynamics is that of the bosonic string\cite{Nielsen}. 
Higher winding number solutions with a single vortex are known as giant vortex 
states in superconductivity, and play an important role in the magnetization
properties of mesoscopic superconductors\cite{Geim,Vital,Peeters,Mallick}.

The above topological argument thus provides the classifying scheme
for all possible vortex solutions (of finite energy) to the 
LGH equations in two (infinite flat) dimensions, in terms of the 
then necessarily integer winding number of the map from the circle 
at planar infinity onto the U(1) gauge group of phase transformations of 
the scalar field $\psi(x)$. Indeed, under both the assumptions of finite 
energy and of an arbitrary collection of discrete zeroes of arbitrary 
positive integer degree in the order parameter $\psi(x)$, it has been 
shown\cite{Taubes} that for a 
specific critical value of the scalar field self-coupling $\lambda_0>0$, 
a critical value which for definiteness we shall refer to as 
$\lambda_0=\lambda_c=\lambda_1$
(the second notation being motivated by the relation to vortices of
integer winding number), all solutions to 
these equations are given in terms of a collection of such magnetic vortices 
each of whose positive or negative integer winding number equals,
in absolute value, the degree of the zero in 
$\psi(x)$ associated to that vortex. Furthermore, still precisely for 
that critical value $\lambda_c$ of the scalar self-coupling, the value for 
the energy of each of these solutions saturates the 
Bomogol'nyi-Prasad-Sommerfeld (BPS) lower bound\cite{Bomo,Prasad}
proportional to its total integer winding number, showing that 
they in fact satisfy first-order differential equations from which their 
second-order equations of motion follow, a tell-tale sign for some 
underlying supersymmetry. 

The stability of these magnetic vortices with respect to fluctuations 
within the same topological classes of configurations has also been studied 
for arbitrary values of the scalar self-coupling\cite{Hindmarsch,Sigal}. 
For a coupling less than 
the critical one, $\lambda_0<\lambda_c$, and a specified integer 
winding number $L$, all such giant vortex states have been shown 
to be stable, thus suggesting that an arbitrary collection of vortices 
of total winding number $L$ would then collapse into a single giant 
vortex with that winding number $L$. In contradistinction, for a coupling 
larger than the critical one, $\lambda_0>\lambda_c$, 
only the $L=\pm 1$ fundamental vortices are stable, 
while giant vortices with $|L|\ge 2$ fall apart into a collection of $|L|$ 
individual fundamental vortices each of winding number ${\rm sign}(L)$. 
Finally specifically at the critical coupling, $\lambda_0=\lambda_c$, 
all configurations corresponding 
to all possible same sign integer partitions of the total winding number $L$ 
are of equal energy, since they all saturate the same BPS lower bound
irrespective of the vortices relative 
positions\cite{Bomo,Prasad,deVega,Rebbi,EWeinberg}.
The physical understanding of these properties stems from a subtle
interplay between, on the one hand, the repulsive (resp. attractive) 
magnetic force acting 
between vortices whose winding numbers are of the same (resp. opposite) 
sign---much like the force between parallel or antiparallel magnetic 
dipoles---and, on the other hand,
the attractive scalar force related to the tendency 
of the system to relax to configurations whose condensate value $|\psi(x)|$
is as close as possible to the constant non vanishing expectation value 
which minimizes the scalar field potential energy (the ratio of the 
latter form of energy density to the former magnetic energy density is 
precisely given by the scalar self-coupling). In particular exactly 
at the critical coupling, $\lambda_0=\lambda_c$, the two types 
of forces balance each other, and vortices do not interact with one another
whatever their relative positions\cite{EWeinberg}. 
Clearly, these very properties also
explain why for Type II superconductors, whose self-coupling is by definition
larger than the critical value, $\lambda_0>\lambda_c$, 
vortices tend to organize themselves
into a triangular lattice of Abrikosov vortices, whose lattice constant
is a function of the scalar self-coupling, as is beautifully confirmed
experimentally\cite{Tinkham,Waldram}.

In spite of the elegant physical insight offered by these results, 
the implicit assumptions on which they rely may not be satisfied by all 
solutions to the LGH equations, including those of finite energy. 
Indeed, it is only in the case of the infinite plane that the restriction to 
configurations of finite energy implies that $|\psi(x)|$ must reach
its vacuum expectation value at infinity in a continuous manner in all
angular directions, and it is this latter fact together with single-valuedness
of $\psi(x)$ which then requires the winding number to be integer.
However, as soon as planar domains of finite spatial extent are considered, 
a situation which is of interest not only to actual superconductors
but clearly also to other systems with phase transitions or topological
states confined to finite spatial volumes, the value of $|\psi(x)|$ on 
the boundary need no longer be the vacuum expectation value, thereby 
allowing {\sl a priori\/} even vanishing values, and thus also possibly 
circumventing the apparent restriction to integer winding numbers only. 
What then becomes of the topological classification above?

Moreover, another assumption always implicit in the results recalled 
above\cite{Taubes} is that the order parameter $\psi(x)$ is such that both
the corresponding function $f(x)$ as defined in (\ref{eq:EQ1}) is positive
everywhere, and that if it does vanish, it does so only at discrete points 
in the plane at which the phase value $\theta(x)$ is then not defined.
However, even though any complex number may always be parametrized
as in (\ref{eq:EQ1}) in terms of its positive amplitude and its phase
defined only modulo $2\pi$, when it comes to a single-valued continuous complex
{\sl function\/} defined on the plane (or any Riemann surface) and 
parametrized as in (\ref{eq:EQ1}), there appears a non trivial correlation 
between the sign of the real function $f(x)$---which may possibly change---and 
the lack of single-valuedness inherent to the phase parametrization 
in terms of $\theta(x)$, this correlation being best highlighted by 
considering the transport of the single-valued order parameter $\psi(x)$ 
along any closed {\sl finite} contour. Clearly, this last remark is topological 
by nature, and thus provides the means to address the question raised above 
with regards to a topological classification of solutions in finite planar 
domains. Note also that this approach takes the contour of the circle 
at planar infinity used in the usual argument, to bring it back at any 
point in the plane at a finite distance, thereby providing a finer-grained 
tool to assess the topological properties of solutions to 
the LGH equations.

As a matter of fact, it was recently pointed out\cite{Gov1} that in finite 
planar domains, beyond the usual vortex configurations, the LGH equations also 
possess annular vortex solutions of finite energy and integer winding
number, whose order parameter $\psi(x)$ vanishes not only at
a given point, but also on a series of concentric closed curves
surrounding that point as well as one another in an almost regular radial 
pattern and at which the function $f(x)$ does indeed alternate in
sign\footnote{The dynamical and thermodynamical stability of these
annular vortices is an open problem.}.
Each such successive annulus is related to a non vanishing closed current
flow always running in the same direction,
responsible for part of the total magnetic flux and whose kinetic 
energy contributes to the total energy of the configuration an almost 
identical amount, thereby leading to an infinite value in the case of
an annular vortex in the infinite plane. Hence, these specific 
properties of such annular vortices explain why they could not be 
uncovered through the general theorems\cite{Taubes}
restricted to only finite energy solutions in the infinite plane and 
such that the function $f(x)$ always remains positive and vanishes only 
at discrete points with some integer degree.

Finally, and this very point lies at the hear of the whole matter, the
above topological argument does not take into account the built-in
U(1) local gauge invariance of the LGH equations which compounds even
further the issue of the lack of single-valuedness of the function
$\theta(x)$ for non zero winding number in correlation with the sign
of the function $f(x)$. In fact, in the same way that for $L\ne 0$
$\theta(x)$ defines a multicovering of the plane as indicated for
instance by its asymptotic value $\theta(x)\simeq -L\phi$, it will
turn out that $f(x)$ must be considered to define a double sheeted
covering of the plane, while nonetheless the order parameter 
$\psi(x)=f(x)e^{i\theta(x)}$ remains continuous, regular and single-valued
throughout spacetime. Thus for example, a single giant vortex with
$L\ne 0$ is associated to a degenerate double covering of the plane
in which the two sheets meet at a single point---namely, the position
of the vortex, at which $f$ varies as $u^{|L|}$, with $u$ measuring
the distance to the vortex axis---, while the double sheeted covering
remains nevertheless regular, continuous and differentiable
everywhere, with the function $f$ having opposite signs on each sheet.

This paper provides a topological classification of solutions
to the LGH equations, which we believe should also prove to be complete since 
it is based on the fine-grained topological consideration mentioned above. 
Clearly, when accounting for the possibility that the real function
$f(x)$ may {\sl a priori\/} take negative as well as positive values
in a continuous fashion, beyond the ordinary and annular vortex 
configurations of integer winding number $L$ there also exist 
vortices of half-integer winding number $L$ such that the order parameter 
function $f(x)$ changes sign an odd number of times when transported 
around some given closed contours surrouding such vortices. More 
specifically, it will be shown that vortices of half-integer 
winding number, which shall be referred to as half-integer vortices for 
short, have the particularity of always being bound onto the edges of an odd
number of two dimensional domain walls (when viewed in three dimensions) 
which end on such vortices, and inside of which the order parameter $\psi(x)$ 
also vanishes on a two dimensional surface. The same type of configuration
may in fact also occur for integer vortices ({\sl i.e.\/} whose winding 
number is integer), in which case the number of such domain walls must
be even. Usual isolated integer vortex configurations are thus a particular
case of the latter type, with a vanishing number of domains walls
ending on each of the vortices. Hence, when viewed in a plane locally 
transverse to any such solution, integer and half-integer vortices 
are characterized by having the order parameter vanish not only at the position 
of the vortex, but also along an even or odd number of continuous lines 
emanating from the vortex itself, in clear contrast again with the 
assumptions of the usual general theorems. Furthermore, in the case of
a finite planar domain, such vortices bound onto the edges of domain walls 
may also be surrounded by an annular pattern of successive closed 
current flows extending up to the boundary, 
in exactly the same manner as already described in the
case of isolated integer vortices. Clearly, when present, such
domain walls are associated to the branch cuts of the double
sheeted covering of the plane which is defined by the function $f(x)$.

Being bound onto the edges of domain walls, such vortex configurations 
also display yet another remarkable property. Even though the order
parameter $\psi(x)$ vanishes inside the domain wall, close to the edge
onto which the vortex is bound a non vanishing closed
current flow must quantum tunnel through the domain wall, in order to
entirely surround the location of the vortex and contribute the required
amount of magnetic flux. In the context of superconductors, this
phenomenon is reminiscent of the Josephson effect for S-I-S 
junctions\cite{Tinkham,Waldram}.
What is a distinctive feature of the integer and half-integer vortices 
bound onto the edges of domain walls however, is that such a quantum 
tunnel current appears within a same single superconductor simply across 
a surface of vanishing order parameter whose location within the material 
could be anywhere and may also vary in time.

Since domain walls are regions of space in which the order parameter
$|\psi(x)|$ takes values which differ from its vacuum expectation value,
they necessarily contribute condensation energy to the system. In first
approximation and for sufficiently extended domain walls, 
this contribution is expected to be proportional to their 
length\footnote{A property reminiscent of the close-to-linear
quark potential in QCD, as S. Vandoren pointed out.}. Indeed, their thickness 
is essentially constant since it is governed by a specific parameter of the 
LGH equations, namely the superconducting coherence length $\xi$, or the 
inverse of the Higgs boson mass $M_h$ in a particle physics parlance 
(see section \ref{Sect2}). Consequently, in the case of spatially bounded
domains, such domain walls may extend all the way to the boundary
of the domain, and lead to finite energy configurations nonetheless.
On the other hand in the case of an infinite spatial extension of the
system, these domain walls must necessarily be of finite extent in
order to lead to solutions of finite energy. In other words, in such a case,
these domain walls must themselves end on some integer or half-integer
vortex within the infinite volume of the sample.

Such considerations thus lead to the following general picture for all
the configurations which solve the LGH equations. The basic entities in terms
of which these solutions may be built up are, on the one hand,
$1/2$-domain walls, and on the 
other hand, annular current flows. By a $1/2$-domain wall, we mean a domain 
wall which ends on a half-integer vortex of winding number $L=1/2$ or 
$L=-1/2$, namely a $1/2$-vortex, but whose other edge may or may not be
bound to another $1/2$-vortex, as the case may be. By an annular current flow, 
we mean a closed current flow carrying no winding number, which may be 
placed around a given vortex but then in combination with other such 
annular flows in the radially outward concentric pattern described previously, 
and which then extends all the way to the boundary of the spatial domain 
(such configurations are thus excluded in the case of an unbounded domain, 
since their energy would be infinite). Indeed, any possible 
vortex configuration of the types described above may be viewed as 
being constructed from these basic building blocks, in exactly the same 
way that until now isolated giant vortices with $|L|\ge 2$ were used to be 
viewed as a collection of $|L|$ individual Abrikosov-Nielsen-Olesen (ANO) 
vortices stacked on top of one another.  Of course, this is also
consistent with the point of view of the double sheeted covering of
the plane advocated in this paper. Such a collection
of isolated giant vortices with $|L|\ge 2$ defines a double covering
in which subsets of the points at which the two sheets meet have coalesced
into single points. In the same spirit, the general solutions including
half-integer vortices and domain walls may thus be viewed as configurations
in which some of these points and coalesced subsets of points have been
split into branch cuts in the double covering of the plane, leading
{\sl in fine\/} to the basic entities considered above.

Thus for example, a $L=3/2$ vortex
may be obtained from three $1/2$-domain walls with $L=1/2$ which share 
a common edge, while the $1/2$-domain walls themselves could or could not 
be stacked on top of one another. Furthermore, each of these $1/2$-domain 
walls could or could not end on another $1/2$-vortex with either $L=1/2$
or $L=-1/2$, but they must do so in the case of a spatial domain of infinite 
extent in order to obtain configurations of finite energy. Indeed, this is the
only constraint which must be imposed on the construction, which implies
that $1/2$-domain walls can never extend up to infinity for unbounded
spatial domains, while they may extend up to the boundary of finite ones.
Since the total winding numbers of such bound $1/2$-domain walls
only take the values $L=0,\pm 1$, the fact that only such 
basic entities are allowed in the case of unbounded spatial domains
is in full agreement with the restriction to only integer total 
winding numbers at spatial infinity as implied by the usual topological 
argument recalled earlier. Nevertheless, the assumptions of the usual 
theorems\cite{Taubes,Sigal} 
are evaded since the order parameter then vanishes not only at discrete 
points in the plane but also along segments of finite length, thereby 
allowing for the existence of these new types of
configurations built up from bound $1/2$-domain walls (annular
current flows are excluded for unbounded domains).
In contradistinction in the case of a planar domain of finite extent,
the total winding number need not be integer, but could also take
half-integer values, since a $1/2$-domain wall may extend all the way 
to the boundary of the planar domain without being bound to another
$1/2$-vortex, leading to a finite energy configuration nonetheless.
Furthermore in this case, such configurations may also be combined
with annular current flows surrouding vortices and extending up to
the boundary of the finite domain. 

The characterization described above of all solutions to 
the LGH equations, which is a direct consequence of the finer-grained
topological analysis based on any coutour at a finite distance in
contrast with the usual argument, also raises the issue of the
stability and lowest energy configuration of bound $1/2$-domain
walls. Given the fact, recalled previously, that magnetic
vortices whose winding numbers are of the same (resp. opposite)
sign repel (resp. attract) each other, while any deviation from 
its vacuum expectation value in the order parameter $\psi(x)$
induces an attractive scalar force, one should expect that
the bound $1/2$-domain wall binding two $1/2$-vortices of opposite
winding numbers $L=1/2$ and $L=-1/2$
would collapse, in its lowest energy state,
to a $L=0$ configuration without any zero in the order parameter.
On the other hand, the same issue in the case of bound $1/2$-domain walls
binding two $1/2$-vortices with the same winding number $L=\pm 1/2$ is far 
less obvious\footnote{Note that such domain walls would repel or attract
one another as well, according to whether the vortices which are bound 
onto their edges have winding numbers of identical or opposite signs. 
For instance in the case of the $L=3/2$ example briefly described previously
and assuming that these bound $1/2$-domain walls do not collapse into
Abrikosov vortices, 
one should expect for this reason that in the infinite domain case
the lowest energy configuration for the $L=3/2$ vortex
is obtained when the three $1/2$-domain walls are not stacked on top of
one another, but rather are spread out into a maximally symmetric
star-like pattern with an angular opening of 120$^{\circ}$ between each, 
the complete state then having a total winding number $L=3$ (see
however section \ref{Subsect3.1}).}.
Given the natural tension built into such vortex and domain wall 
states, their energy must increase with their curvature or bending, 
so that the lowest energy configurations are expected to be straight
vortices and domain walls. Nevertheless, as pointed out above, for 
sufficiently large separation the energy of such states must also 
grow essentially linearly with the distance between the two 
$1/2$-vortices, $1/2$-domains behaving very much like rubber bands
in such a regime because of the condensation energy stored in the
domain wall. Hence, it is only when the two $1/2$-vortices have
a non vanishing overlap that the competition between the repulsive and
attractive forces acting on them could be such that the equilibrium
lowest energy configuration would be reached for some specific non vanishing
separation rather than a fully collapsed state. 
The possible overlap between vortices depends on their
size, which is governed by yet another parameter of the LGH equations, 
namely the Meissner penetration length $\lambda$, or the inverse of 
the gauge boson mass $M_\gamma$ in a particle physics parlance (see 
section \ref{Sect2}). In fact, the ratio of this latter length scale
to that which determines the width of domain walls is directly related
to the value of the scalar self-coupling. Consequently, one would expect
that there might exist some critical value $\lambda_{1/2}$
of the scalar self-coupling marking the boundary between, on the one hand,
unstable bound $1/2$-domain walls which would 
always collapse to $L=\pm 1$ ANO vortices whatever 
the separation between the bound $1/2$-vortices, and, on the other hand,
stable bound $1/2$-domain 
walls which would stabilize into configurations such that the domain wall 
is streched to some finite length. In fact, given the situation which 
applies to the stability of integer vortices against fluctuations 
which are also of integer winding number\cite{Sigal}, 
one would expect that it is for values of the
scalar self-coupling $\lambda_0$ larger than this critical value 
$\lambda_{1/2}$ that the latter occurrence would be realized.
Furthermore for the same reason, it may even be possible\footnote{However,
one should not overlook the fact that the stability analysis\cite{Sigal} 
of isolated integer vortices considered only fluctuations within the
same topology class, thereby not necessarily accounting for the
possibility of half-integer vortices nor domain walls.} that the
two critical values $\lambda_1=\lambda_c$ and $\lambda_{1/2}$, 
thus characterizing the stability properties of isolated
integer vortices and of half-integer vortices bound
by domain walls, should be equal to one another, and thus be equal to the 
value $\lambda_c$ for which the usual isolated integer vortices also saturate 
the BPS bounds.

Such considerations thus also raise the possibility that the $L=\pm 1$ 
ANO vortex could be unstable against splitting into 
two $L=\pm 1/2$ vortices bound onto the edges of a bound $1/2$-domain wall
with $L=\pm 1$,
and which would stabilize from one another at some specific non vanishing 
distance whose value would depend on that of the scalar self-coupling.
Even though some arguments and indicative results will be presented 
concerning this issue, no conclusion can be drawn at this stage 
unfortunately, since it requires a full-fledged dedicated numerical analysis 
which still needs to be completed. Obviously, this specific point, and in 
particular the value for the possible equilibrium distance as a function of 
the scalar self-coupling, is of great interest, for instance, to the physical 
properties of magnetic vortices in superconductors. For example, given a
value for that coupling such that the ANO vortex would be split,
the Abrikosov lattice acquires a fine-grained structure
in which the centers of all bound $1/2$-domain walls still define the usual
triangular lattice, but with all individual bound $1/2$-domain walls
then all parallel to one another and aligned with one of the three
symmetry axes of the fundamental triangular lattice cell, as suggested 
by symmetry and minimal energy considerations.

The above analysis, based on the fine-grained topological argument
involving closed contours at any finite distance, has thus unravelled
quite an unexpectedly rich classification of vortex, domain wall
and annular current flow configurations (the latter only in bounded
spatial domains) of integer and half-integer winding numbers,
solving the LGH equations and extending
the well established isolated vortex solutions. Such states can
all be built out of some basic entities bound and stacked together,
namely the $1/2$-domain wall and the annular current flow in the case of
bounded spatial domains, and the bound $1/2$-domain walls in unbounded spatial
domains. Even though the bound $1/2$-domain walls of winding numbers $L=\pm 1$ 
may not be stable against collapse into the $L=\pm 1$ 
ANO vortices for some regimes of scalar self-coupling,
nevertheless the mere existence of these fluctuation modes of bound
$1/2$-domain walls with winding numbers $L=0,\pm 1$
solving the LGH equations, must have important consequences with regards to the 
dynamical properties of systems described by the LGH equations when coupled 
to external sources of electromagnetic disturbances, the first 
example in point being again obviously superconductors.

The integer and half-integer winding number configurations bound by
domain walls considered in this paper
are also cha\-rac\-te\-ri\-zed by specific cut structures in the functions
$f(x)$ and $\theta(x)$, and a double sheeted covering of the plane
or a finite domain of it
(see section \ref{Subsect2.4}). Such cuts however, have no bearing 
whatsoever on the physical consistency of these configurations, since not
only does the order parameter $\psi(x)$ remain single-valued, regular
and continuous
throughout spacetime, but also it is only $|\psi(x)|^2=f^2(x)$ which is a
gauge invariant physical observable, measuring in the context of
superconductivity for example, the local (relative) density of Cooper
pairs, and in the particle physics context, the local condensation
of the scalar field. As will become clearer in the course of the
analysis, it is the U(1) gauge invariance properties of the system which 
provide the essential guarantee for the physical consistency and existence 
of half-integer winding number and domain wall solutions to the LGH equations.

In recent years, similar half-integer and fractional winding number
topological con\-fi\-gu\-ra\-tions have in fact also been discussed within the 
framework of models other than the abelian U(1) Higgs model, which could 
possibly have applications in high temperature superconductors and 
superfluids\cite{Volovik}, in the QCD confinement problem\cite{Cornwall}, 
and for so-called Alice strings\cite{Schwarz}
in models beyond the Standard Model of particle physics. 
What distinguishes all these other instances of fractional winding number 
from the half-integer vortices discussed in this paper, is that these 
other models possess some spontaneously broken {\sl non abelian\/} internal 
symmetry, whether global or locally gauged, such that when taken 
around a closed contour the matter fields of these systems end up pointing 
in a different direction in the representation space of their internal 
symmetry. What is remarkable with the present half-integer vortices is that 
a similar mechanism is at work in spite of having only an abelian U(1) 
gauged symmetry, by carefully keeping track, when taken around closed 
contours, of the possible change of sign in $f(x)$ as well as its lack of 
single-valuedness in correlation with that of the phase parameter $\theta(x)$,
while the order parameter $\psi(x)$ itself remains nonetheless
single-valued throughout space. As will become clear hereafter, this
particular feature is achieved through the U(1) gauge invariance properties
of the considered system and the correlation which exists between its
gauge dependent variables.

The paper is organized as follows. In the next section, the general
equations of the abelian U(1) Higgs model will be considered,
to lead to the topological classification of integer and half-integer
vortices. Section \ref{Sect3} then particularizes to the
two dimensional case, namely the plane transverse to such solutions,
to discuss specific limits and properties related to the possible BPS 
bounds these solutions could saturate. Single half-vortex solutions in 
finite disks and annuli are then considered in section \ref{Sect4},
including the results of some preliminary numerical simulations. 
Similar issues are then addressed in section \ref{Sect5}
in the case of the bound $1/2$-domain wall,
including some preliminary indicative numerical results as to the possible 
instability of the $L=1$ vortex. Finally, open issues and further outlook 
onto possible consequences and applications of these 
topological configurations of half-integer winding number of the abelian 
U(1) Higgs model are discussed in the Conclusions.

\section{The Topological Analysis}
\label{Sect2}

\subsection{The LGH equations}
\label{Subsect2.1}

In particle physics units, such that $\hbar=1$, $c=1$, $\epsilon_0=1$
and $\mu_0=1/(\epsilon_0c^2)=1$---$\epsilon_0$ and $\mu_0$ being the
vacuum electric and magnetic permitivities, respectively---the abelian 
U(1) Higgs model is defined by the following action\footnote{Note that
the specific case of a four dimensional spacetime is considered, but that
the results of this paper may easily be extended to higher dimensions
and thereby lead to domain wall solutions of integer and half-integer
winding numbers of arbitrary spacelike dimensionalities and bound to the edge
of domain walls of one more spacelike dimension. Such configurations
are very much reminiscent of the D-branes\cite{Polchinski} 
in string and M-theory (see also
section \ref{Subsect3.3}), thus raising many further issues of 
interest which are beyond the scope of this work.
In particular, in the same way that the low energy effective 
dynamics of the ANO vortex is described by the Nambu-Goto
action\cite{Nielsen}, 
the same dynamics for such domain walls and those considered in 
this paper is necessarily also provided by the Nambu-Goto action extended to
world-volumes of the appropriate dimension.
In the same vein, one could also consider possible
generalizations to some complex $p$-form coupled to a real $q$-form 
with a spontaneously broken local symmetry.},
\begin{equation}
\int_{(\infty)}d^4x^\mu\,\left\{-\frac{1}{4}F^{\mu\nu}F_{\mu\nu}\,+\,
\mid(\nabla_\mu+iqA_\mu)\psi\mid^2-\frac{1}{4}\lambda_0
\left(\mid\psi\mid^2-a^2\right)^2\right\},
\end{equation}
with $F_{\mu\nu}=\nabla_\mu A_\nu-\nabla_\nu A_\mu$ and 
$\nabla_\mu\equiv\partial/\partial x^\mu$ ($\mu=0,1,2,3$; the notation 
$\partial_\mu$ is reserved for another purpose hereafter), while our choice 
of Minkowski metric signature is $(+---)$. Here, $q$ stands for the
U(1) charge or gauge coupling of the complex scalar field $\psi(x)$, and 
$a>0$ for its expectation value with the units of a mass scale which is 
then also the dimension of the field $\psi(x)$ itself.

By construction, this system possesses a U(1) local gauge invariance under,
\begin{equation}
\psi'(x)=e^{i\chi(x)}\psi(x)\ \ \ ,\ \ \ 
A'_\mu(x)=A_\mu(x)-\frac{1}{q}\nabla_\mu\chi(x),
\end{equation}
$\chi(x)$ being an arbitrary function which is regular throughout 
spacetime. This symmetry is spontaneously broken through the Higgs potential 
of dimensionless scalar field self-coupling $\lambda_0>0$, thereby leading to 
a massive gauge boson $\gamma$ and a massive Higgs scalar $h$ both of zero
U(1) charge such that
\begin{equation}
M_\gamma=|qa|\ \ \ ,\ \ \ M_h=\sqrt{2\lambda_0}\,a.
\end{equation}

As is already implicit in most of the discussion presented in the
Introduction, we shall rather work with another choice of units,
directly appropriate to an application to superconductors. Nevertheless,
a translation from one choice of units to the other is straightforward
enough, for which the rules are given hereafter. Normalizing
the scalar field $\psi$---representing the Cooper pair wave function---to
its vacuum expectation value---namely the square-root of the BCS
condensate in the bulk in the absence of any magnetic field---, the
corresponding action is given by\cite{Gov1}
\begin{equation}
\epsilon_0c^2\int dt\int_{(\infty)}d^3\vec{x}\left\{
-\frac{1}{4}F^{\mu\nu}F_{\mu\nu}+\frac{1}{2}
\left(\frac{\Phi_0}{2\pi\lambda}\right)^2
\left[\mid(\nabla_\mu+i\frac{q}{\hbar}A_\mu)\psi\mid^2-
\frac{1}{2\xi^2}\left(\mid\psi\mid^2-1\right)^2\right]\right\},
\end{equation}
where now all quantities are expressed in S.I. units, with of course
$x^\mu=(x^0,\vec{x})$ and $x^0=ct$, $A^\mu=(\Phi/c,\vec{A})$---$\Phi$ 
being the electromagnetic scalar potential---and 
$F_{\mu\nu}=\nabla_\mu A_\nu-\nabla_\nu A_\mu$. 
In the above expression, $\Phi_0$ stands for the unit of
quantum of flux
\begin{equation}
\Phi_0=\frac{2\pi\hbar}{|q|}>0,
\end{equation}
$q=-2e<0$ being of course the Cooper pair electric charge, while
$\lambda$ and $\xi$ are the temperature dependent magnetic (and electric)
penetration and coherence lengths, respectively. Note how these two length 
scales indeed measure the relative contributions to the effective action 
above, and thus also to the associated energy, of the magnetic and condensate
energy densities. Indeed, the scale $\xi$ weighs the 
contribution of the gauge covariantized gradient of any deviation in
$\psi$ from its vacuum expectation value against its potential
energy, while this total condensation energy in turn is measured
against the electromagnetic energy contribution through the scale
$\lambda$. Note also that one has,
\begin{equation}
\frac{E^i}{c}=-\frac{\partial}{\partial x^i}\frac{\Phi}{c}-
\frac{1}{c}\frac{\partial}{\partial t}A^i=F^{i0}\ \ ,\ \ 
B^i=\epsilon^{ijk}\left[\frac{\partial}{\partial x^j}A^k-
\frac{\partial}{\partial x^k}A^j\right]=-\epsilon^{ijk}F^{jk}\ \ ,\ \ 
i,j,k=1,2,3,
\end{equation}
where $\vec{E}$ and $\vec{B}$ stand of course for the electric and
magnetic fields, respectively, and $\epsilon^{ijk}$ is the totally
antisymmetric Levi-Civita tensor in three dimensions with $\epsilon^{123}=+1$.

Comparing now the above two actions, the translation between the
particle physics units and the ``superconductor" ones is as follows,
\begin{equation}
a\longleftrightarrow\frac{\Phi_0}{2\pi\lambda}\ \ ,\ \
\lambda_0 a^2\longleftrightarrow\frac{1}{\xi^2}\ \ ,\ \
M_\gamma\longleftrightarrow\frac{1}{\lambda}\ \ ,\ \
M_h\longleftrightarrow\frac{\sqrt{2}}{\xi}\ \ ,\ \
\frac{M_h}{M_\gamma}\longleftrightarrow\sqrt{2}\kappa,
\end{equation}
where $\kappa=\lambda/\xi$ defines the Landau-Ginzburg (LG) parameter
of the superconductor. In particular, the critical value $\lambda_c=\lambda_1$
for the scalar self-coupling mentioned in the Introduction then corresponds
to the value $\kappa=\kappa_c=\kappa_1$ with $\kappa_c=1/\sqrt{2}$, which,
in the particle physics context, thus implies $M_h=M_\gamma$ corresponding
to $\lambda_0=\lambda_c=\lambda_1=1$.

In terms of the ``superconductor units", the LGH equations then read
as follows. For the LGH equation proper, we have the covariantized
Landau-Ginzburg equation
\begin{equation}
\left[\nabla_\mu+i\frac{q}{\hbar}A_\mu\right]
\left[\nabla^\mu+i\frac{q}{\hbar}A^\mu\right]\psi+
\frac{1}{\xi^2}\left(|\psi|^2-1\right)\psi=0,
\end{equation}
which is thus coupled to the inhomogeneous Maxwell 
equations\footnote{The homogeneous Maxwell equations are of course a 
consequence of the definition of the field strength $F_{\mu\nu}$ in terms of
the gauge potential $A_\mu$, being the corresponding Bianchi identities.}
\begin{equation}
\nabla^\nu F_{\nu\mu}=\mu_0 J_{{\rm em}\,\mu},
\end{equation}
where the conserved electromagnetic current density
$J^\mu_{\rm em}=\left(c\rho_{\rm em},\vec{J}_{\rm em}\right)$, such that
$\nabla_\mu J^\mu_{\rm em}=0$, is given by
($\psi^*$ being the complex conjugate of $\psi$)
\begin{equation}
\mu_0J_{{\rm em}\,\mu}=\frac{i\hbar}{2q\lambda^2}\left[
\psi^*\left(\nabla_\mu\psi+i\frac{q}{\hbar}A_\mu\psi\right)-
\left(\nabla_\mu\psi^*-i\frac{q}{\hbar}A_\mu\psi^*\right)\psi\right].
\end{equation}

As a matter of fact, still another choice of normalized units is far
more convenient, using the canonical length scale provided by $\lambda$ 
and the canonical magnetic field value provided by $\Phi_0/(2\pi\lambda^2)$.
Let us thus introduce the following normalized spacetime 
coordinates\cite{Gov1,Gov2}
\begin{equation}
\tau=\frac{x^0}{\lambda}=\frac{ct}{\lambda}\ \ ,\ \ 
\vec{u}=\frac{\vec{x}}{\lambda},
\end{equation}
such that
\begin{equation}
\nabla_\mu=\frac{1}{\lambda}\partial_\mu\ \ ,\ \ 
\partial_\mu=(\partial_0,\vec{\partial})\ \ ,\ \ 
\partial_0=\partial_\tau\equiv\frac{\partial}{\partial\tau}\ \ ,\ \ 
\vec{\partial}\equiv\frac{\partial}{\partial\vec{u}},
\end{equation}
while in the electromagnetic sector, the normalized quantities
$\vec{e}$, $\vec{b}$, $f_{\mu\nu}$, $a^\mu=(\varphi,\vec{a})$ 
and $J^\mu=(J^0,\vec{J}\,)$ are defined as follows,
\begin{equation}
\vec{B}=\frac{\Phi_0}{2\pi\lambda^2}\,\vec{b}\ \ ,\ \ 
\frac{\vec{E}}{c}=\frac{\Phi_0}{2\pi\lambda^2}\,\vec{e}\ \ ,\ \ 
F_{\mu\nu}=\frac{\Phi_0}{2\pi\lambda^2}\,f_{\mu\nu}\ \ ,\ \ 
A^0=\frac{\Phi_0}{2\pi\lambda}\,\varphi\ \ ,\ \ 
\vec{A}=\frac{\Phi_0}{2\pi\lambda}\,\vec{a}, 
\end{equation}
\begin{equation}
\mu_0c\rho_{\rm em}=\frac{1}{\lambda}\frac{\Phi_0}{2\pi\lambda^2}\,J^0\ \ ,\ \ 
\mu_0\vec{J}_{\rm em}=\frac{1}{\lambda}\frac{\Phi_0}{2\pi\lambda^2}\,\vec{J}. 
\end{equation}

In terms of these normalized variables, the LGH equation then reads
\begin{equation}
\left(\partial_\mu-ia_\mu\right)\left(\partial^\mu-ia^\mu\right)\psi=
\kappa^2\left(1-\mid\psi\mid^2\right)\psi\ \ {\rm or}\ \
\left[(\partial_\tau-i\varphi)^2-(\vec{\partial}+i\vec{a})^2\right]\psi=
\kappa^2\left(1-|\psi|^2\right)\,\psi,
\label{eq:LGcomplex}
\end{equation}
while the inhomogeneous Maxwell equations are
\begin{equation}
\partial^\nu f_{\nu\mu}=J_\mu\ \ {\rm or}\ \
\vec{\partial}\cdot\vec{e}=J^0\ \ ,\ \ 
\vec{\partial}\times\vec{b}-\partial_\tau\vec{e}=\vec{J},
\end{equation}
with the components of the Lorentz covariant electromagnetic current 
density $J^\mu=(J^0,\vec{J}\,)$ given by
\begin{equation}
J_\mu=-\frac{1}{2}i\left[\psi^*(\partial_\mu\psi-ia_\mu\psi)-
(\partial_\mu\psi^*+ia_\mu\psi^*)\psi\right],
\end{equation}
or
\begin{equation}
J^0=-\frac{1}{2}i\left[\psi^*(\partial_\tau\psi-i\varphi\psi)-
(\partial_\tau\psi^*+i\varphi\psi^*)\psi\right]\ ,\ 
\vec{J}=\frac{1}{2}i\left[\psi^*(\vec{\partial}\psi+i\vec{a}\psi)-
(\vec{\partial}\psi^*-i\vec{a}\psi^*)\psi\right],
\end{equation}
and satisfying the conservation equation,
\begin{equation}
\partial_\mu J^\mu=0\ \ {\rm or}\ \
\partial_\tau J^0+\vec{\partial}\cdot\vec{J}=0,
\end{equation}
which as always follows from the inhomogeneous Maxwell equations.
Moreover, we then also have
\begin{equation}
f_{\mu\nu}=\partial_\mu a_\nu-\partial_\nu a_\mu\ \ {\rm or}\ \
\vec{e}=-\vec{\partial}\varphi-\partial_\tau\vec{a}\ \ ,\ \ 
\vec{b}=\vec{\partial}\times\vec{a},
\end{equation}
relations which imply the homogeneous Maxwell equations as the corresponding
Bianchi identities,
\begin{equation}
\epsilon^{\mu\nu\rho\sigma}\partial_\nu f_{\rho\sigma}=0\ \ {\rm or}\ \
\vec{\partial}\times\vec{e}+\partial_\tau\vec{b}=\vec{0}\ \ ,\ \ 
\vec{\partial}\cdot\vec{b}=0,
\label{eq:Maxwellhomo}
\end{equation}
$\epsilon^{\mu\nu\rho\sigma}$ being the four dimensional Levi-Civita
totally antisymmetric tensor with $\epsilon^{0123}=+1$.

Let us now also introduce the Lorentz covariant current $j^\mu=(j^0,\vec{j})$
whose components are defined by dividing the current density $(-J^\mu)$ 
opposite to the electromagnetic one by the square $|\psi|^2$ of the spacetime
local condensate value $|\psi|$,
\begin{equation}
j_\mu=-\frac{J_\mu}{|\psi|^2}=\frac{1}{2}i\partial_\mu
\ln\left(\frac{\psi}{\psi^*}\right)+a_\mu,
\label{eq:littlej1}
\end{equation}
or
\begin{equation}
j^0=-\frac{J^0}{|\psi|^2}=\frac{1}{2}i\partial_\tau
\ln\left(\frac{\psi}{\psi^*}\right)+\varphi\ \ ,\ \ 
\vec{j}=-\frac{\vec{J}}{|\psi|^2}=-\frac{1}{2}i\vec{\partial}
\ln\left(\frac{\psi}{\psi^*}\right)+\vec{a}.
\label{eq:littlej2}
\end{equation}
In particular, one then finds\cite{Gov2}
\begin{equation}
f_{\mu\nu}=\partial_\mu j_\nu-\partial_\nu j_\mu\ \ {\rm or}\ \
\vec{e}=-\partial_\tau\vec{j}-\vec{\partial}j^0\ \ ,\ \ 
\vec{b}=\vec{\partial}\times\vec{j}.
\label{eq:London}
\end{equation}
Note that the last of these relations is nothing but the
celebrated second London equation, which explains the Meissner
effect in superconductors. On the other hand, the second of these
relations is the appropriate Lorentz covariant extension of
the celebrated first London equation. Indeed,
the abelian U(1) Higgs model provides an effective theory of
superconductivity which is also Lorentz covariant, and in which
electric and magnetic fields play roles dual to one another under Lorentz
boosts. Contrary to the usual LG and London equations which are not
Lorentz covariant, and do not allow for electric fields in superconductors,
the covariant formalism must necessarily imply new phenomena in
specific regimes where electric fields can no longer be ignored.
Clearly a covariant formalism implies specific differences
in the time dependent dynamics of superconductors, but it does so
already for static configurations of applied electric and magnetic
fields, as discussed recently in Ref.\cite{Gov2}.

Finally, let us give the expression for the free energy of the system
in our choice of normalized units,
when subjected to some external electric and magnetic fields
$\vec{e}_{\rm ext}$ and $\vec{b}_{\rm ext}$. One has
\begin{equation}
\begin{array}{r l}
E=\frac{\lambda^3}{2\mu_0}\left(\frac{\Phi_0}{2\pi\lambda^2}\right)^2
\int_{(\infty)}d^3\vec{u}&\left\{
\left[\vec{e}-\vec{e}_{\rm ext}\right]^2+
\left[\vec{b}-\vec{b}_{\rm ext}\right]^2+\right.\\
 & \\
& \left.+\mid(\partial_\tau-i\varphi)\psi\mid^2+
\mid(\vec{\partial}+i\vec{a})\psi\mid^2+
\frac{1}{2}\kappa^2\left(1-\mid\psi\mid^2\right)^2
-\frac{1}{2}\kappa^2\right\}.
\end{array}
\end{equation}
Note that by having subtracted in the integrand
the constant term in $\kappa^2/2$,
we have chosen to fix the zero value of the free energy at the
normal-superconducting phase transition for which $\psi=0$
and $\vec{e}=\vec{e}_{\rm ext}$, $\vec{b}=\vec{b}_{\rm ext}$.
Depending on the consideration we wish to emphasize, especially
when discussing solutions in the infinite plane, sometimes we shall
not include that constant term in the evaluation of the total energy
of the system.

The physical interpretation of the overall constant factor multiplying
this expression for the free energy should be clear, since it
corresponds to the magnetic energy of a constant magnetic field
of value $\Phi_0/(2\pi\lambda^2)$ over a volume $\lambda^3$, precisely
the two scales with respect to which all quantities have been
normalized.

It is also useful to revert to particle physics units given our choice
of normalized quantities. From the translation rules given previously, 
it follows that distance scales are measured in units of $1/M_\gamma$,
magnetic and electric fields in units of $M^2_\gamma/|q|$,
and finally the energy in units of
\begin{equation}
\frac{\lambda^3}{2\mu_0}\left(\frac{\Phi_0}{2\pi\lambda^2}\right)^2
\longleftrightarrow\frac{M_\gamma}{2q^2},
\end{equation}
thus displaying the well-known fact that solitonic solutions
to field theory equations have mass values which are proportional 
to the gauge boson mass scale and inversely
proportional to the squared gauge coupling constant.

\subsection{The polar parametrization}
\label{Subsect2.2}

Given the different expressions above, let us now consider the
polar decomposition of the scalar field,
\begin{equation}
\psi(x)=f(x)\,e^{i\theta(x)},
\end{equation}
where $f(x)$ is thus a real function and $\theta(x)$ is an angular
variable defined modulo $\pi$ rather than $2\pi$. Indeed, since the sign
of the real function $f(x)$ cannot be constrained to be always positive,
which, for reasons discussed in the Introduction, would be an unjustifiable 
restriction in the case of a complex function as opposed to a complex number, 
the arbitrariness in the overall sign of $f(x)$ translates into a definition 
of $\theta(x)$ only modulo $\pi$ (the reason for this arbitrariness will be
fully understood in section \ref{Subsect2.4}, on basis of the double
sheeted covering of the plane mentioned in the Introduction). 
Moreover, the lack of single-valuedness 
in this quantity, which is thus correlated to the choice of an overall sign 
for the real function $f(x)$, is also compounded with the U(1) local 
gauge freedom inherent to the system, an issue which shall be addressed 
more closely at the end of this section and in the next one.

The conserved electromagnetic current density $J^\mu$, the source term for the
inhomogeneous Maxwell equations, is then given by
\begin{equation}
J_\mu=-f^2j_\mu\ \ {\rm or}\ \
J^0=-f^2\,j^0\ \ ,\ \ \vec{J}=-f^2\,\vec{j}\ \ ,
\end{equation}
with thus the conservation equation
\begin{equation}
\partial_\mu\left(f^2j^\mu\right)=0\ \ {\rm or}\ \
\partial_\tau\left(f^2j^0\right)+\vec{\partial}\cdot\left(f^2\vec{j}\right)=0.
\label{eq:conservation}
\end{equation}
In particular, the phase variable $\theta$ is determined from the
equations (\ref{eq:littlej1}) and (\ref{eq:littlej2}), namely,
\begin{equation}
\partial_\mu\theta=-j_\mu+a_\mu\ \ {\rm or}\ \
\partial_\tau\theta=-j^0+\varphi\ \ ,\ \ 
\vec{\partial}\theta=\vec{j}-\vec{a}.
\label{eq:theta1}
\end{equation}
Separating the real and imaginary parts of the LGH equation 
(\ref{eq:LGcomplex}), one finds that for $f\ne 0$
the imaginary part is equivalent to 
the current conservation equation as befits a U(1) gauge invariant theory, 
while the real part reads,
\begin{equation}
\left[\vec{\partial}^2-\partial_\tau^2\right]f=
\left(\vec{j}^2-{j^0}^2\right)f-\kappa^2(1-f^2)f.
\label{eq:LG}
\end{equation}

In order to identify an organizational principle within this set of
coupled differential equations, let us now substitute the covariant
London equations (\ref{eq:London}) into the inhomogeneous Maxwell
equations. This leads to,
\begin{equation}
\left[\vec{\partial}^2-\partial^2_\tau\right]j^0=
f^2j^0-\partial_\tau\left(\partial_\tau j^0+\vec{\partial}\cdot\vec{j}\right)
\ \ ,\ \
\left[\vec{\partial}^2-\partial^2_\tau\right]\vec{j}=f^2\vec{j}+
\vec{\partial}\left(\partial_\tau j^0+\vec{\partial}\cdot\vec{j}\right).
\label{eq:j}
\end{equation}
The expected appearance of the Klein-Gordon operator 
$\partial_\mu\partial^\mu=[\partial^2_\tau-\vec{\partial}^2]$ in these 
different equations suggests to also establish a similar equation for 
$\theta$, given the relations in (\ref{eq:theta1}), leading to
\begin{equation}\left[\vec{\partial}^2-\partial^2_\tau\right]\theta=
\left(\partial_\tau j^0+\vec{\partial}\cdot\vec{j}\right)-
\left(\partial_\tau\varphi+\vec{\partial}\cdot\vec{a}\right).
\label{eq:theta2}
\end{equation}

In conclusion, we have thus obtained the following set of equations,
which may be expressed either in a manifestly covariant form or in terms
of each of their components separately. The Lorentz covariant time dependent
LGH equation is thus
\begin{equation}
\partial_\mu\partial^\mu\,f=j_\mu j^\mu f+\kappa^2(1-f^2)f,
\end{equation}
or equivalently the equation given in (\ref{eq:LG}).
This LGH equation is coupled to the inhomogeneous Maxwell
equations\footnote{These equations may also be expressed in first-order
form as $\partial^\nu f_{\nu\mu}=-f^2j_\mu$,
provided one uses the relations between the electric and magnetic
fields and the current $j^\mu=(j^0,\vec{j})$ which are 
defined by the covariant London equations (\ref{eq:London}).}
\begin{equation}
\partial_\nu\partial^\nu j_\mu=-f^2j_\mu+
\partial_\mu\left(\partial_\nu j^\nu\right),
\end{equation}
or in component form the equations in (\ref{eq:j}),
from which the current conservation equation (\ref{eq:conservation}) follows.
Once this coupled set of equations for $f$ and $j^\mu=(j^0,\vec{j})$
solved (for a given choice of boundary conditions to be discussed presently), 
the associated electric and magnetic fields within the superconductor 
are determined from the Lorentz covariant London equations (\ref{eq:London}),
\begin{equation}
f_{\mu\nu}=\partial_\mu j_\nu-\partial_\nu j_\mu\ \ {\rm or}\ \
\vec{e}=-\partial_\tau\vec{j}-\vec{\partial}j^0\ \ ,\ \ 
\vec{b}=\vec{\partial}\times\vec{j}, 
\label{eq:London2}
\end{equation}
from which the homogeneous Maxwell equations (\ref{eq:Maxwellhomo}) follow.
One then needs to identify gauge potentials $a^\mu=(\varphi,\vec{a})$ 
associated to these two fields, such that 
\begin{equation}
f_{\mu\nu}=\partial_\mu a_\nu-\partial_\nu a_\mu\ \ {\rm or}\ \
\vec{e}=-\vec{\partial}\varphi-\partial_\tau\vec{a}\ \ ,\ \ 
\vec{b}=\vec{\partial}\times\vec{a}, 
\end{equation}
to also determine the phase variable $\theta$ through the equations
given in (\ref{eq:theta1}), or the second-order equation in 
(\ref{eq:theta2}), namely
\begin{equation}
\partial_\mu\theta=-j_\mu+a_\mu\ \ ,\ \ 
\partial_\mu\partial^\mu\,\theta=-\partial_\mu j^\mu+\partial_\mu a^\mu.
\label{eq:theta3}
\end{equation}
Finally, the free energy of the system
may also be expressed solely in terms of the variables $f$, $j^0$
and $\vec{j}$, the electric and magnetic fields within the superconductor
being obtained from the covariant London equations (\ref{eq:London2}),
\begin{equation}
\begin{array}{r l}
E=\frac{\lambda^3}{2\mu_0}\left(\frac{\Phi_0}{2\pi\lambda^2}\right)^2
\int_{(\infty)}d^3\vec{u}&\left\{
\left[\vec{e}-\vec{e}_{\rm ext}\right]^2+
\left[\vec{b}-\vec{b}_{\rm ext}\right]^2+\right.\\
 &  \\
&  +\left.
\left(\partial_\tau f\right)^2+\left(\vec{\partial}f\right)^2+
\left({j^0}^2+\vec{j}^2\right)f^2+\frac{1}{2}\kappa^2\left(1-f^2\right)^2
-\frac{1}{2}\kappa^2\right\}.
\end{array}
\label{eq:freeen}
\end{equation}

The above relations thus determine the system of coupled equations
which is to be solved, once the relevant boundary conditions specified.
Outside the superconductor, the scalar field $\psi$ vanishes of course,
and since the covariant London equations no longer apply,
the relevant complete set of inhomogeneous and homogeneous Maxwell 
equations then has to be considered on its own, including the possibility 
of applied fields or sources for these fields providing further boundary 
conditions. At the boundary with the superconductor, 
matching conditions for the electromagnetic field strength $f_{\mu\nu}$ and 
potential $a_\mu$ must be enforced. Moreover, when considering variations
of the action or the free energy of the system with respect to the 
order parameter $\psi$, one concludes that the current 
$(\vec{\partial}+i\vec{a})\psi$, namely the gauge covariantized space 
gradient of $\psi$, must have a vanishing normal component at those 
sections of the superconductor boundary which are in contact with an 
isolating material. Separating the real and imaginary parts of that 
quantity, this condition thus requires that both the electromagnetic 
current $\vec{J}=-f^2\vec{j}$ as well as the gradient $\vec{\partial}f$ must 
have a vanishing component normal to such boundaries. These different
conditions would thus appear to provide the complete set of
necessary boundary conditions, with one exception however, to be discussed
in the next section. Indeed, the integrated London equations,
through the associated electromagnetic flux values, provide further
constraints of a global character which encode the topological vortex
structure within the superconductor, which the local form of the
covariant London equations in (\ref{eq:London2}) cannot account for.
These ``global boundary conditions", are the last essential conditions 
which should define
unique and gauge invariant solutions to the LGH and Maxwell equations above.

Until now, we have been careful to maintain manifest the Lorentz covariance
properties of the system throughout. It should be clear
that in the limit of time independent configurations, namely static
or stationary ones, the above set of equations reduces to the familiar
Landau-Ginzburg equations for superconductors solely submitted to magnetic
fields (thus with $\vec{e}=\vec{0}$, $\varphi=0$ and $j^0=0$). 
However, also for reasons advocated in Ref.\cite{Gov2}, when con\-si\-de\-ring
time dependent configurations or situations involving electric fields
as well, the usual equations can no longer be applicable, since in
particular the usual first London equation, $\partial_\tau\vec{j}=-\vec{e}$
(rather than the covariant one above, 
$\partial_\tau \vec{j}+\vec{\partial}j^0=-\vec{e}$),
implies that superconductors cannot sustain electric fields in static
configurations, in clear contradiction with Lorentz covariance. 
Another way to realize that the
usual LG equations cannot be Lorentz covariant is to consider the
non relativistic limit in which the parameter $c$ is taken to infinity.
Since in S.I. units electric fields are measured in 
units of $\vec{E}/c$ relative to magnetic fields $\vec{B}$, in that limit
both any time as well as any electric field dependencies decouple from 
the above equations, which then also reduce again to the usual time 
independent non covariant LG equations. It is thus of great interest 
to explore the 
physical consequences in relativistic and electric regimes of the abelian U(1)
Higgs model as the canonical Lorentz covariant extension of the usual
Landau-Ginzburg effective field theory description of 
superconductivity\cite{Weinberg}.
Note however that the solutions which are described in this paper
are static ones, in the absence of any electric field, and thus
solve in fact also the usual non covariant LG equations coupled to
Maxwell's equations for magnetic fields, even though these states
in fact define Lorentz covariant configurations solving the LGH equations 
when subjected to arbitrary Lorentz boosts acting from their rest frame.

The advantage of having separated the system of coupled equations through
the organization just described is not only that first one needs
to solve only for the quantities $f$, $j^0$ and $\vec{j}$, and only then
for the phase variable $\theta$ through the knowledge of the electric and
magnetic fields and their associated gauge potentials\footnote{As will
be discussed in the next section, one should keep in mind however, 
the subtle connection between these two sets of variables which involves 
the vortex topological structure through the integrated London equations.}. 
More importantly perhaps, for obvious physics reasons
the quantities $f^2$, $f^2j^0$ and $f^2\vec{j}$ are in fact gauge
invariant variables, directly amenable to physical observation, at least
in principle, which is a property also shared by the electric and 
magnetic fields $\vec{e}$ and $\vec{b}$ only. Indeed, U(1) gauge 
transformations of the field variables are given by,
\begin{equation}
\psi'=e^{i\chi}\psi\ \ ,\ \ A'_\mu=A_\mu-\frac{\hbar}{q}\nabla_\mu\chi
\ \ ,\ \ a'_\mu=a_\mu+\partial_\mu\chi,
\end{equation}
so that
\begin{equation}
\theta'=\theta+\chi\ \ ,\ \ \varphi'=\varphi+\partial_\tau\chi\ \ ,\ \ 
\vec{a}'=\vec{a}-\vec{\partial}\chi,
\label{eq:gaugetransf}
\end{equation}
where $\chi(\tau,\vec{u})$ is thus an arbitrary local function which is
regular throughout spacetime. As was discussed above, the real function $f$
is defined only up to an overall sign whose choice, however, leaves all the
above equations invariant as it should, and which indeed does not lead to
any physical consequence. Nevertheless, this choice of overall sign
in $f$ is also correlated to a constant shift by $\pm\pi$ in $\theta$,
again without consequences for physical observables. This correlated
arbitrariness in both $f$ and $\theta$ is then compounded further in
the case of $\theta$ because of the U(1) gauge
symmetries of the system, which induce arbitrary spacetime dependent
shifts in that variable in correlation with specific transformations
in the gauge potential $a_\mu$. In particular, note that the equations
(\ref{eq:theta1}) or (\ref{eq:theta2}) and (\ref{eq:theta3})
display this correlation explicitly through their manifest gauge invariance.

Consequently, except for the choice of overall sign in the function
$f$, which is without physical significance, all the gauge dependency of the
system resides in the variables $\theta$, $\varphi$ and $\vec{a}$.
The remaining variables, namely $f$---except for its overall sign---,
$j^0$, $\vec{j}$, $\vec{b}$ and $\vec{e}$ are all gauge invariant
quantities, whose determination through the above equations is thus
decoupled from that of the gauge dependent variables.
Hence the advantage in using the procedure outlined above for the
resolution of the coupled system of equations. In fact, as will be discussed
in the next section, the U(1) local gauge freedom is such that
there always exists a choice of gauge
for which the function $\theta$ is then uniquely determined in terms of a 
specific non regular function $\theta_0$ which completely encodes the entire 
topological vortex structure of a given configuration, and in terms of 
which the gauge potential $a_\mu$ is then uniquely constructed from
(\ref{eq:theta3}) as $a_\mu=\partial_\mu\theta_0+j_\mu$.
Hence indeed, through an appropriate choice of gauge
associated to a specific vortex configuration in spacetime,
only the LGH and inhomogeneous Maxwell equations (\ref{eq:LG}) 
and (\ref{eq:j}) remain to be solved for the quantities $f$ and 
$j^\mu=(j^0,\vec{j})$ subjected to the appropriate boundary conditions.

To conclude, let us also make the important following point.
Even though the currents $j^0$ and $\vec{j}$ are gauge invariant,
they are not directly amenable to physical observation, since they are so
only through the electromagnetic current density $(-f^2j^0,-f^2\vec{j})$.
Hence, it is only at those points in spacetime where $f$ does not vanish
that the current $(j^0,\vec{j})$ is well-defined, while at those
locations where $f$ does vanish, it may well be that this current possesses
singularities just mild enough to be screened by the vanishing order
parameter $\psi$. As we shall see, this is precisely what happens
not only at the location of vortices, but also on the domain walls
ending on them. In fact, the expressions in (\ref{eq:littlej1})
and (\ref{eq:littlej2}) are already indicative of the logarithmic cut 
structures that may appear in $j^0$, $\vec{j}$, $f$ and $\theta$ at the 
zeroes of the order parameter $\psi$ in a way which depends on its winding 
number at these points. Note also that the same remark applies
to the singular character of the gauge dependent potential $a_\mu$ at the 
location of vortices, a well known property typical of topological 
configurations
related to gauge fields which is made explicit in the present instance
through the appearance of the singular function $\theta_0$ in
the determination of $a_\mu$, while electric and magnetic fields 
are nonetheless well defined throughout spacetime, including at the 
location of vortices.

\subsection{Configurations of integer and half-integer winding number}
\label{Subsect2.3}

Let us now apply the fine-grained topological analysis described
in the Introduction to the general framework which has been set-up
so far. For this purpose, let us consider any closed contour
$C$ in spacetime and measure the total electromagnetic flux through
any two dimensional surface $S$ with $C$ as a boundary. When normalized
to the quantum of flux $\Phi_0$, this electromagnetic flux is
given by the following expression for the associated surface
integral (the notation should not be confused with that
for the electromagnetic scalar potential in S.I. units),
\begin{equation}
\Phi[C]=-\frac{1}{2\pi}\int_S d^2u^{\mu\nu}\,f_{\mu\nu}.
\end{equation}

If the contour $C$ and the surface $S$ are purely spacelike,
this quantity $\Phi[C]$ does indeed measure, in units of $\Phi_0$, the
total magnetic flux through that surface. In the case of a purely
timelike rectangular contour with fixed boundaries in space, the
quantity $\Phi[C]$ is in fact a measure of the electrostatic potential
difference between those two points in space. The purpose of
this general analysis is to make manifest once again the Lorentz
covariance properties of the discussion, which will eventually be
restricted to time independent configurations in a planar geometry only. 

In fact, by its very definition, the flux $\Phi[C]$ is not only a 
Lorentz scalar, but also a U(1) gauge invariant quantity. Indeed,
through Stokes theorem, we also have
\begin{equation}
\Phi[C]=-\frac{1}{2\pi}\oint_C du^\mu\,a_\mu,
\end{equation}
which is recognized to be precisely the U(1) gauge invariant Wilson loop
associated to the contour $C$. Using now the general first-order relation 
in (\ref{eq:theta3}), we obtain
\begin{equation}
\Phi[C]=-\frac{1}{2\pi}\oint_C du^\mu\,\partial_\mu\theta\ -\
\frac{1}{2\pi}\oint_C du^\mu\,j_\mu=
-\frac{1}{2\pi}\oint_C du^\mu\,\partial_\mu\theta\ +\
\frac{1}{2\pi}\oint_C du^\mu\,\frac{1}{f^2}J_\mu .
\end{equation}
Note that this relation, when considered for all possible contours $C$
in spacetime, provides the integrated global expression of the 
covariant London equations (\ref{eq:London2}) which in their local
differential form cannot encode the topological information which we are 
about to cha\-rac\-te\-ri\-ze. Working in terms of this global integrated 
form of the
covariant London equations is crucial to the resolution of the LGH
equations in a way which also accounts for the possible topological
vortex structure of its solutions. The values $\Phi[C]$ for all possible
contours $C$ in spacetime thus also provide
the ``global boundary conditions" connecting the gauge invariant and 
gauge dependent variables and their semi-decoupled system of
differential equations in the manner described in the previous section.

Since the angular variable $\theta(\tau,\vec{u})$ need not
be single-valued throughout spacetime, the contour integral of its 
spacetime gradient $\partial_\mu\theta$ could {\sl a priori\/}
lead to an arbitrary real winding number $L[C]$ in correspondence with the 
specific contour $C$ being considered, such that
\begin{equation}
\frac{1}{2\pi}\oint_C du^\mu\,\partial_\mu\theta=-L[C].
\label{eq:LC}
\end{equation}
Hence for an arbitrary contour $C$, the integrated (covariant) London 
equations read,
\begin{equation}
\Phi[C]=L[C]-\frac{1}{2\pi}\oint_C du^\mu\,j_\mu=
L[C]+\frac{1}{2\pi}\oint_C du^\mu\frac{1}{f^2}J_\mu .
\end{equation}
In the case of a contour $C$ which only lies in space directions,
this general expression reduces to the following relation
for the corresponding magnetic flux,
\begin{equation}
\Phi[C]=L[C]+\frac{1}{2\pi}\oint_C d\vec{u}\cdot\vec{j}=
L[C]-\frac{1}{2\pi}\oint_C d\vec{u}\cdot\frac{\vec{J}}{f^2} .
\end{equation}
When considering a contour $C$ at infinity in the case of an unbounded
superconducting domain, 
this expression leads to the well known property of
magnetic flux quantization at infinity,
since the electromagnetic current density contribution then vanishes
while $L[C]$ must be an integer for the topological reason recalled
in the Introduction.

In fact for all choices of contours, including those at a finite distance,
the values for $L[C]$ are restricted by the necessary
single-valuedness of the order parameter $\psi$, which thus constrains
the lack of single-valuedness in the phase parameter $\theta$ in
correlation with the fact that the function $f$ is real but with
a sign which could vary throughout spacetime, while its overall sign is left
unspecified. As explained in the Introduction, to best consider that issue
let us now imagine taking the scalar field $\psi$ around the closed
contour $C$. Given the winding number $L[C]$ associated to this
contour as defined in (\ref{eq:LC}), under transport around $C$ the 
phase variable $\theta$ is shifted by the quantity $(-2\pi L[C])$,
\begin{equation}
\theta\stackrel{C}{\longrightarrow}\theta-2\pi L[C]\ \ ,\ \ 
e^{i\theta}\stackrel{C}{\longrightarrow}e^{-2i\pi L[C]}\,e^{i\theta}.
\end{equation}
In turn, since the order parameter $\psi$---which is assumed not to be
vanishing throughout spacetime---must be single-valued under such 
a transformation, the function $f$ itself must then transform according to
\begin{equation}
\psi\stackrel{C}{\longrightarrow}\psi\ \ ,\ \ 
f\stackrel{C}{\longrightarrow} e^{2i\pi L[C]}f.
\end{equation}
However, since the function $f$ is to take real values only (but not
necessarily positive ones only), consistency of such transformations 
thus restricts the possible winding numbers $L[C]$ for any 
contour $C$ to integer as well as half-integer values only. In the case 
of integer winding number values, the function
$f$ recovers its original sign after transport around the cor\-res\-pon\-ding
closed contours, while the phase $\theta$ is then shifted by an even multiple
of $\pi$. In the case of half-integer values however, $f$ changes
sign under such a transformation, in correlation with the shift by an
odd multiple of $\pi$ in the phase $\theta$. Nevertheless in either case,
the complex scalar field $\psi$ remains single-valued throughout
spacetime, as it should. Note also that if one artificially restricts
the function $f$ to take positive values only, as is done in the usual
classification theorems\cite{Taubes}, only integer winding number configurations
survive the analysis with the possibility of a vanishing order parameter
at discrete locations only but not in a continuous fashion such
as inside some domain walls.

Given this topological characterization of solutions, 
in the case of a half-integer winding number $L[C]$ the function $f$ must
necessarily vanish and change sign an odd number of times when taken
around the closed contour $C$, while in the case of an integer winding
number $f$ may vanish and change sign an even number of times, including
zero of course.
Moreover in the case of a non vanishing winding number $L[C]$---be it integer
or half-integer---for a contour
$C$ which shrinks to a point, necessarily the order parameter
$\psi$ and thus the function $f$ must vanish at that point, since the phase
variable $\theta$ is then ill-defined at that point being shifted by 
$(-2\pi L[C])$ when taken around that point. Hence by continuity, given a 
contour $C$ of non vanishing winding number $L[C]$ and a surface 
$S$ with $C$ as a boundary, the general solutions to the LGH equations
may be characterized by having the order parameter $\psi$ vanish at some 
point on $S$ as well as on a series of continuous lines lying within $S$ 
and emanating from that point, the number of such lines being even in the 
case of an integer winding number and odd for a half-integer one. Viewed 
in three space dimensions, such configurations correspond to vortex 
solutions extending along some spacelike one dimensional
curve and to which an even or odd 
number of two dimensional finite or semi-infinite domain walls are 
attached at one of their edges according to whether the value of the vortex
winding number is integer or half-integer, respectively\footnote{Note 
that in the general case, the winding number 
$L[C]$ is a function of time for any specific fixed contour $C$ in 
space. This function is piece-wise constant, with integer or 
half-integer discontinuities at those specific instants when integer or 
half-integer vortices leave or enter the closed contour $C$.}. 

In order to characterize the possible network of vortices and domain
walls specific to a given configuration, let us turn again to 
the gauge invariance properties of the abelian U(1) Higgs model.
As emphasized previously, the flux $\Phi[C]$ is both Lorentz and
gauge invariant. In fact, under gauge transformations, see
(\ref{eq:gaugetransf}), the winding number contribution to $\Phi[C]$
transforms as,
\begin{equation}
L'[C]=-\frac{1}{2\pi}\oint_C du^\mu\,\partial_\mu\theta'=
-\frac{1}{2\pi}\oint_C du^\mu\,\partial_\mu\left[\theta+\chi\right]
=L[C]-\frac{1}{2\pi}\oint_C du^\mu\,\partial_\mu\chi,
\end{equation}
while the contribution from the current $j_\mu$ is gauge invariant
by itself. Hence, it would appear that the winding numbers of any
solution could be shifted away through some appropriate gauge transformation
$\chi$ whose contour integrals exactly cancel the winding number contribution
to the flux $\Phi[C]$ for all contours $C$. 
Such gauge transformations however, leading to
topology change in the gauge and Higgs fields, are not allowed. Indeed,
they would necessarily correspond to functions $\chi$ which are not
well defined throughout spacetime, possessing singularities precisely
canceling those of the vortex solutions. Since only gauge transformations
$\chi$ which are regular and single-valued throughout spacetime
are acceptable, the winding numbers $L[C]$ as well as the fluxes
$\Phi[C]$ are indeed gauge invariant
physical observables for whatever choice of contour $C$.

On the other hand, for solutions of non vanishing winding number,
the phase variable $\theta$ must necessarily possess
specific singularities at some points in spacetime, since some of its
contour integrals are non vanishing even when shrinking to a point
which then cor\-res\-ponds to the location of a vortex. Nevertheless, the 
regular component of $\theta$ may be changed at will through arbitrary gauge
transformations (\ref{eq:gaugetransf}) parametrized by regular functions
$\chi$, without changing the winding numbers of $\theta$ but affecting
nonetheless the gauge potential $a_\mu$ accordingly. This remark thus
suggests that it may be possible to gauge away entirely any regular
contribution to the phase variable $\theta$ of the order parameter $\psi$,
leaving over only the singular contributions responsible for all winding
numbers encoded in $\theta$ while modifying the gauge potential
appropriately through the relevant gauge transformation $\chi$.
To show how this is can be achieved, let us assume to have constructed
a specific function $\theta_0$ whose winding numbers for all spacetime
contours $C$ reproduce exactly all those of $\theta$ given a specific
vortex configuration solving the LGH equations
(such a function will be constructed explicitly in the next section).
Since the difference $(\theta_0-\theta)$ is necessarily a regular function
which is well defined throughout spacetime, let us then apply the gauge 
transformation of parameter $\chi=\theta_0-\theta$, leading to
\begin{equation}
\theta'=\theta_0\ \ ,\ \ a'_\mu=a_\mu-\partial_\mu\theta+\partial_\mu\theta_0.
\end{equation}
However, given the second-order equation in (\ref{eq:theta3}), the
transformed gauge potential $a'_\mu$ is such that
\begin{equation}
\partial^\mu a'_\mu=\partial_\mu a^\mu-\partial_\mu\partial^\mu\theta
+\partial_\mu\partial^\mu\theta_0=
\partial_\mu j^\mu+\partial_\mu\partial^\mu\theta_0.
\end{equation}
Finally, since the current $j^\mu$ is gauge invariant, this last 
relation for $a'_\mu$ remains invariant under all further 
gauge transformations $\chi$ such 
that $\partial_\mu\partial^\mu\chi=0$. Whether only the trivial solution
$\chi=0$ obeys this equation depends on the choice of boundary conditions
for the electromagnetic sector of the system at infinity\footnote{We
refrain from being more specific on this point since one often considers
the application of homogeneous external electric and magnetic fields outside
the superconductor.}. Whatever the case may be in this respect,
the important conclusion is thus as follows. Given a specific
configuration of bound vortices and domain walls whose topological
structure is encoded into the function $\theta_0$, one may impose
the ``vortex gauge fixing condition"\footnote{Note that this gauge fixing 
condition is Lorentz invariant, and that for time independent axially 
symmetric configurations, it reduces to the Coulomb-London gauge fixing 
condition $\vec{\partial}\cdot\vec{a}=0$ (see the next section).},
\begin{equation}
\partial_\mu a^\mu=\partial_\mu j^\mu+\partial_\mu\partial^\mu\theta_0.
\label{eq:gaugefixing}
\end{equation}
The solution for the gauge dependent variables of the system is then
given by
\begin{equation}
\theta=\theta_0\ \ ,\ \ a_\mu=\partial_\mu\theta_0+j_\mu .
\end{equation}
Depending on the choice of boundary conditions at infinity, the vortex
gauge fixing (\ref{eq:gaugefixing})
may not be complete, in which case both $\theta$ and
$a_\mu$ are defined only up to those gauge transformations whose parameter
function $\chi$ also obeys the equation
\begin{equation}
\partial_\mu\partial^\mu\chi=0,
\label{eq:Laplace}
\end{equation}
but leading nevertheless to a construction which is consistent with the 
above expressions for the solutions for both $\theta$ and $a_\mu$,
since both these quantities are then gauge transformed accordingly.
Hence, the vortex gauge fixing condition (\ref{eq:gaugefixing})
together with the knowledge of the function $\theta_0$ does indeed
completely solve the LGH equations in the sector of gauge dependent
variables, as announced previously. Note that when non trivial gauge
transformations such that $\partial_\mu\partial^\mu\chi=0$ exist,
the arbitrariness that such a situation leads to in terms of $\theta$
and $a_\mu$ may in fact be absorbed into the choice of function
$\theta_0$ which then also suffers the same physically irrelevant ambiguity.

\subsection{The function $\theta_0$ and the double sheeted covering
of the plane}
\label{Subsect2.4}

In order to understand how to construct the function $\theta_0$
in the general situation, let us first restrict to time independent
configurations in the planar case. In other words, we assume that
the system has been dimensionally reduced to two flat space dimensions,
with the time independent fields dependent only on the two
coordinates of that plane, the magnetic field $\vec{b}$ 
purely transverse to that plane and the electric field $\vec{e}$ and
charge distribution $j^0$
vanishing. Choosing coordinates on the plane---normalized to the length 
scale $\lambda$---which are either cartesian,
with $x=u^1$, $y=u^2$, or polar, with $u=\sqrt{x^2+y^2}$,
$\phi={\rm Arctan}(y/x)$ and the specific evaluation $-\pi\le\phi\le+\pi$,
it is useful to introduce the complex combinations
\begin{equation}
z=x+iy=ue^{i\phi}\ \ ,\ \ z^*=x-iy=ue^{-i\phi}.
\end{equation}

Let us now consider a specific collection of $K$ vortices described as follows.
Each of these vortices, labelled by $k=1,2,\dots,K$, has integer or
half-integer winding number $L_k=N_k/2$, $N_k$ being thus an even or
odd integer, respectively. The total winding number of the configuration
is thus $L=N/2$ with $L=\sum_{k=1}^KL_k$ and $N=\sum_{k=1}^KN_k$.
In addition, each of these vortices has a position
in the plane corresponding to the complex parameter
$z_k=x_k+iy_k=u_ke^{i\phi_k}$.
In order to construct the function $\theta_0$ associated to this vortex 
configuration, let us also introduce for each value of $k=1,2,\dots,K$
a collection of $N_k$ complex functions $\theta_{k,n_k}(z)$
labelled by $n_k=1,2,\dots,N_k$ and dependent only on the complex variable $z$.
Then, a choice of function $\theta_0(x,y)=\theta_0(u,\phi)$ which
reproduces all the contour integrals associated to this vortex
configuration is given by the following expression,
\begin{equation}
\theta_0=\frac{1}{4}i\sum_{k=1}^K\sum_{n_k=1}^{N_k}
\ln\left[\frac{e^{-i\theta_{k,n_k}(z)}\left(z-z_k\right)}
{e^{+i\theta^*_{k,n_k}(z^*)}\left(z^*-z^*_k\right)}\right],
\end{equation}
where a specific evaluation of the complex logarithmic function is to
be chosen (for example with its branch cut along the real negative axis
in the plane, namely at $\phi=\pm\pi$ for $\ln(z)$).

The meaning of this construction is as follows. The introduction of the
index $n_k$ taking a total of $N$ values is in fact related to the
interpretation discussed in the Introduction which views all possible
configurations as built up from $1/2$-vortices and $1/2$-domain walls
which are combined and stacked on top of one another. Each value for the
index $n_k$ thus refers to each such $1/2$-vortex as a basic building block,
with in particular the corresponding function $\theta_{k,n_k}(z)$
specifying the position in the plane $(x,y)$ of the logarithmic
branch cut in the quantum phase $\theta=\theta_0$ which is associated to that 
particular $1/2$-vortex and whose branch point is at the position of 
that vortex. However, note that one may also write, up to some specific 
integer multiple of $\pi$ to be added on the r.h.s.,
\begin{equation}
\theta_0=\frac{1}{2}i\sum_{k=1}^KL_k\ln
\left[\frac{z-z_k}{z^*-z^*_k}\right]\ +\
\frac{1}{4}\sum_{k=1}^K\sum_{n_k=1}^{N_k}
\left[\theta_{k,n_k}(z)+\theta^*_{k,n_k}(z^*)\right].
\end{equation}
Obviously, since the double sum on $k$ and $n_k$ in the r.h.s.
of this expression defines a real function $\chi(x,y)$ which trivially
satisfies the equation $\partial_\mu\partial^\mu\chi=0$, this term
may be gauged away altogether, leaving over only the first simple sum over $k$.
In other words, the positions of the different logarithmic branch cuts in the
function $\theta=\theta_0$ may be moved around in the plane $(x,y)$ through
arbitrary gauge transformations; only their branch points are fixed at
the $1/2$-vortices locations. In particular, these branch cuts may be chosen
to lie exactly  along the lines of vanishing order parameter within
the domain walls that may exist. Hence, when taking that degree of freedom
into account which allows to gauge away all functions $\theta_{k,n_k}(z)$, 
there is no loss of generality in the following specific choice, for
instance, for the function $\theta_0$ associated
to such an arbitrary vortex configuration,
\begin{equation}
\theta_0=\frac{1}{2}i\sum_{k=1}^K L_k
\ln\left[\frac{z-z_k}{z^*-z^*_k}\right]\ \ ,\ \ 
e^{i\theta_0}=\prod_{k=1}^K\left[
\frac{z-z_k}{z^*-z^*_k}\right]^{-L_k/2}.
\label{eq:theta0}
\end{equation}

As a particular example, consider a single vortex of winding number $L$
at the center of the plane. One then has
\begin{equation}
e^{i\theta_0}=e^{-iL\phi},
\end{equation}
so that the order parameter takes the general form
\begin{equation}
\psi(u,\phi)=f(u,\phi)\,e^{-iL\phi}.
\end{equation}
Having chosen to work with the angular range $-\pi\le\phi\le+\pi$, clearly
the phase factor $\theta=\theta_0=-L\phi$ of this configuration has a branch 
cut at $\phi=\pm\pi$ starting at the origin, while $f(u,\phi)$ must then
change sign on that branch cut for a half-integer winding number $L$.
A similar discussion applies in general, by considering the neighbourhood
of any vortex in the plane. Hence, whenever half-integer vortices
are present---which, as discussed in the Introduction, should indeed be 
the generic situation---, these branch cut properties show that the 
function $f$ is in fact defined over a double covering of the plane,
or a finite domain of it in the case of a bounded superconductor,
with branch points at the positions of these vortices and branch cuts
on those lines on which the order parameter vanishes within the domain
walls which are bound to these vortices. The function $f(u,\phi)$ then changes
sign on these branch cuts, in such a way that it be continuous on the
specific double sheeted covering of the plane which is associated to the 
considered
configuration of bound vortices and domain walls. Clearly, the same
general picture also applies to integer vortices bound onto domain walls.
It is only in the case of isolated integer vortices, thus not bound onto
the edges of any domain walls---namely, the only situation thought to be 
possible until now---, that the function $f(u,\phi)$ is in fact continuously
single-valued on the plane $(x,y)$ itself, rather than on some double
covering of it, in which case it is indeed justified to assume that
it be always positive (or negative) everywhere, with the exception of
isolated points where it vanishes. Nonetheless, given the arbitrary
overall sign for the function $f(u,\phi)$, it remains more appropriate,
even in such a case, to still view that function as being defined on
a double sheeted covering of the $(x,y)$ plane in which the branch cuts
between the two sheets have degenerated in a continuous fashion into
single points which then correspond to the positions of the isolated
integer vortices. 

Hence by allowing all the degrees of
freedom hidden in the polar parametrization of the order parameter,
$\psi=fe^{i\theta}$, to manifest themselves in a way made consistent
by the U(1) gauge invariance of the system, we see that the general
solution to the LGH equations is associated to a specific double covering
of the plane or a finite domain of it, 
in the manner just described. In particular, the unspecified
overall sign of the function $f$ is then also seen to correspond to a choice
of sheet in this double sheeted covering of the plane.
When only isolated integer vortices are involved, these two sheets
become almost disconnected by touching only at isolated points,
thereby allowing the sign of $f$ to be fixed to
remain either positive (or negative) as has always been 
assumed implicitly until now.
But when domain walls are involved as well, then necessarily the two
sheets become intertwined in a topologically non trivial manner precisely
along these domain walls. Finally, in the case
of a planar domain of finite extent, some of the branch cuts in $f$ 
may extend all the way up to the boundary of the domain, and
this general picture is compounded
even further by the possibility of annular current flows surrouding
the vortices, and thus adding further concentric closed branch cut structures
in an almost periodic fashion to the double covering of the planar domain,
with the function $f$ changing sign at the boundaries of successive
annular flows.

As a last remark concerning the function $\theta_0$ as constructed above
for time independent planar configurations,
note that it satisfies the equation
$\partial_\mu\partial^\mu\theta_0=0$. Hence in such a situation the
vortex gauge fixing condition (\ref{eq:gaugefixing}) reduces to the
condition that $\partial_\mu a^\mu=\partial_\mu j^\mu$, namely
$\partial_x a^1+\partial_y a^2=\partial_x j^1+\partial_y j^2$.
For an axially symmetric time independent configuration, we then have
the Coulomb-London gauge fixing condition $\vec{\partial}\cdot\vec{a}=0$.

Let us now return to the general situation of a collection of integer
and half-integer vortices bound onto the edges of a collection of
domain walls, all in a time dependent fashion and moving in a three
dimensional space. In such a case, it is possible to define\footnote{Such
a coordinate system is in fact not unique.} a curvilinear
system of coordinates associated to a specific foliation of spacetime,
such that two of the coordinates define spacelike surfaces which are locally
transverse to each of the vortices throughout space, with the third spacelike
coordinate locally transverse to these surfaces, and finally the fourth
timelike coordinate transerve---in terms of the Minkowski metric---to
all three spacelike coordinates. In other words, at each instant in time
it is possible to view the vortex system as being obtained from
the previous planar description, by appropriately
bending these planes, and by stacking
them on top of one another in a transverse direction in such a manner
that all vortices and domain walls remain continuous throughout space.
The coordinates in the planes locally transverse to all vortices then
play the same role as the cartesian or polar coordinates considered
in the planar case above. Hence, it is possible
to construct again the function $\theta_0$ in this general
case using exactly the same functional expression as above, but this time 
in terms of the curvilinear coordinates associated to these transverse planes.
The only difference is then that the complex coordinates $z_k(\tau)$ and
functions $\theta_{k,n_k}(z,\tau)$ specifying the positions of the vortices
and of the logarithmic branch cuts of $\theta=\theta_0$, become functions of 
time as well (with the functions $\theta_{k,n_k}(z,\tau)$ amenable to being
gauged away as before). This is thus how the topological structure and
time dependency of a given vortex and domain wall configuration may be
encoded into the function $\theta_0$, from which the solutions for
$\theta=\theta_0$ and $a_\mu=\partial_\mu\theta_0+j_\mu$ may be obtained 
in the vortex gauge
$\partial_\mu a^\mu=\partial_\mu j^\mu+\partial_\mu\partial^\mu\theta_0$
in the manner described previously.
Note that in this general case, the function $\theta_0$
no longer obeys the equation $\partial_\mu\partial^\mu\theta_0=0$.

This concludes the general discussion of the topological classification
of all solutions to the coupled Maxwell and LGH equations.
The sector of gauge dependent variables, namely $\theta$ and $a_\mu$,
is determined in terms of a function $\theta_0$, possibly defined up
to some gauge transformation, which encodes the gauge invariant topological
structure characterized through all the winding numbers which are
associated to a given configuration of vortices and domain
walls. The sector of gauge invariant variables, namely $f$---up to an overall
sign---, $j^0$ and $\vec{j}$, is determined by solving the coupled
LGH and Maxwell equations (\ref{eq:LG}) and (\ref{eq:j}), subjected
to specific boundary conditions which include the ``global boundary
conditions" obtained from the electromagnetic flux values $\Phi[C]$\
for all possible closed contours $C$ in spacetime, which in fact provide
the global integrated London equations and depend on the vortex
configuration through its winding numbers. This last set of equations and
boun\-da\-ry conditions is totally gauge invariant, which is an
advantage when solving this system. Furthermore,
one should keep in mind that the
function $f$ is in fact defined on a double sheeted covering of the plane,
with branch points at vortices and branch cuts along the surfaces of 
vanishing order parameter lying within the domain walls.
Finally, the (free) energy of such configurations may be obtained
from the general expression (\ref{eq:freeen}).

In this latter respect, it is interesting to consider a few sample
situations corresponding to a single straight vortex of winding number
$L$ placed at the center of the infinite plane, for which we have seen that
the winding phase factor is given by $e^{-iL\phi}$. 
In the case of an isolated 
integer vortex, the physical configuration is in fact invariant under 
rotations around the vortex, since indeed only the order parameter
$\psi(u,\phi)=f(u)e^{-iL\phi}$ is not cylindrically symmetric because
of its winding phase factor, while no domain wall is attached to the
vortex which would otherwise break this rotational invariance.
Consequently, at a fixed distance $u$ from the vortex, the order
parameter retains a fixed vacuum expectation value $|f(u)|<1$, so that
when taking the order parameter around the contour of radius $u$
from $\phi=-\pi$ to $\phi=+\pi$, the values reached by $\psi$ in the
complex plane all lie on the circle of radius $|f(u)|<1$ which is then
wound $L$ times starting from $\psi=e^{i\pi L}f(u)=(-1)^Lf(u)$ 
and ending back at the same point
on that circle. This circle also lies somewhere in between the
absolute minimum $|\psi|=1$ and the local maximum $\psi=0$ of the
surface in the shape of the bottom of a wine bottle which is defined by the
Higgs potential $\left(1-|\psi|^2\right)^2$.

Consider now again an integer vortex, say with $L=1$, but this time
bound to an even number of domain walls. Rotational invariance being
then broken, when going around the contour of radius $u$ the
order parameter $\psi(u,\phi)=f(u,\phi)e^{-iL\phi}$
no longer retains a constant norm. Rather,
starting from one of the domains walls where $\psi=0$, $f=0$, the order
parameter moves in a continuous fashion from the top of the Higgs potential
down to some maximal value for the norm $|\psi|<1$ and back to the local
maximum at $\psi=0$ by following a closed path which passes through the latter
point as often as there are domain walls bound to the vortex.
More specifically, when there are two domain walls, a single such loop
is followed from top to bottom and back twice in the same direction.
But when there are four domain walls for instance, then 
the entire closed circuit lying on the surface of the Higgs potential
has the shape of a four-leaved clover whose center is at the top of
the Higgs potential, and so on for still larger
even numbers of domain walls.

Similarly in the case of $L=1/2$ and a single domain wall, the order
parameter $\psi(u,\phi)=f(u,\phi)e^{-iL\phi}$ follows only once 
a closed loop running from the top to some lowest
point on the Higgs potential surface and back. When there are three
domain walls bound to the $L=1/2$ vortex, a three-leaved clover
shape is obtained, and so on. Taking a $L=3/2$ vortex to which three
domain walls are bound, again a single loop running from top to bottom
and back is followed three times, while when only one domain wall is bound to
the $L=3/2$ vortex the closed circuit is followed only once and reaches
the top of the Higgs potential only once, but it then also explores 
surroundings lying on opposite sides of the top of the Higgs potential
by following a spiraling path which leaves and returns at the top in a
direction perpendicular to that of its lowest point.

In fact, for symmetry and energy reasons, one would expect that
for half-integer vortices of winding number $L=N/2$
the configuration of lowest energy would either be that with all
$|N|$ bound domain walls stacked on top of one another, or else
spread out in a maximally symmetric star-like shape with an angular 
opening of $2\pi/|N|$ between each. 
For a given vortex of integer winding number $L=N/2$, there also
exists the third possibility that no domain wall whatsoever is
bound to the vortex.
Presumably, which of these different possibilities is actually realized 
in each of these cases would also
depend on the value of the scalar self-coupling $\kappa$ or $\lambda_0$.
Indeed, it is for either one of these numbers of domain walls that the order 
parameter, when taken around a closed contour surrounding the vortex,
would explore the more regions closer to the bottom of the Higgs potential,
thus leading to a smaller condensation energy contribution.
Only a detailed numerical analysis however, would resolve this issue.

\section{Two Dimensions and BPS Bounds}
\label{Sect3}

\subsection{Two dimensional reduction}
\label{Subsect3.1}

For the remainder of the paper, we shall restrict the discussion to time
independent con\-fi\-gu\-ra\-tions in the plane, in the manner already
described previously. Furthermore, since when considering bounded
domains in the plane these will have the topology of either a disk or
an annulus, the appropriate choice of coordinates is the polar
one with the $u$ and $-\pi\le\phi\le+\pi$ variables also introduced
previously. In particular, a disk topology will have a radius $b$,
namely $u_b=b/\lambda$ given our choice of normalized units, while an
annular topology will have inner and outer (normalized) radii
$u_a=a/\lambda$ and $u_b>u_a$, respectively.

An arbitrary time independent configuration in the plane is represented
in terms of the following set of quantities.
The electric field $\vec{e}$, scalar gauge potential $\varphi$ and
charge distribution $j^0$ all vanish. 
The magnetic field $\vec{b}$ is perpendicular to the plane 
with a single component $b(u,\phi)$, such that
$\vec{b}(u,\phi)=\left(0,0,b(u,\phi)\right)$.
Finally, the vectors $\vec{j}(u,\phi)$ and
$\vec{a}(u,\phi)$ only have radial and azimuthal components, namely
$\vec{j}(u,\phi)=\left(j_u(u,\phi),j_\phi(u,\phi),0\right)$ and 
$\vec{a}(u,\phi)=\left(a_u(u,\phi),a_\phi(u,\phi),0\right)$. 
All these non vanishing quantities are
functions of $u$ and $\phi$ only, but neither of the coordinate
transverse to the plane nor of time, as are then also the order
parameter $\psi(u,\phi)$ and its polar decomposition variables $f(u,\phi)$
and $\theta(u,\phi)$. In particular, note that this restriction on the
decomposition of the vector potential $\vec{a}$ is consistent with the 
vortex gauge fixing condition (\ref{eq:gaugefixing}).
As a matter of fact, it also proves useful to introduce the following
notation,
\begin{equation}
j(u,\phi)\equiv j_u(u,\phi)\ \ ,\ \ 
g(u,\phi)\equiv uj_\phi(u,\phi),
\end{equation}
since these quantities appear naturally in all expressions.

Given such specific symmetry restrictions---namely translation
invariance in time and in the direction transverse to the plane,
or in other words parallel to the straight vortices and domain walls---, it
is immediate to determine the form of the ensuing system of equations.
For the LGH equation (\ref{eq:LG}), one finds
\begin{equation}
\left[\,u\partial_uu\partial_u\,+\,\partial^2_\phi\,\right]f=
\left[\left(uj\right)^2+g^2\right]f-
\kappa^2u^2\left(1-f^2\right)f,
\label{eq:LG1}
\end{equation}
while the inhomogeneous Maxwell equations (\ref{eq:j}) reduce to
(in first-order form)
\begin{equation}
u\partial_ub=f^2g\ \ ,\ \ 
\partial_\phi b=-f^2uj, 
\label{eq:MAX2}
\end{equation}
where the magnetic field (transverse component) is given by the local 
London equation
\begin{equation}
b=\frac{1}{u}\partial_ug-\frac{1}{u}\partial_\phi j.
\label{eq:L2}
\end{equation}
The current conservation equation, which follows from (\ref{eq:MAX2})
of course, reads,
\begin{equation}
u\partial_u\left(f^2uj\right)+\partial_\phi\left(f^2g\right)=0.
\end{equation}

The set of coupled differential equations (\ref{eq:LG1}) and 
(\ref{eq:MAX2}) is subjected to a series of boun\-da\-ry
conditions. First, the ``global boundary conditions", corresponding
to the global London equations, are
\begin{equation}
\Phi[C]=L[C]+\frac{1}{2\pi}\oint_Cd\vec{u}\cdot\vec{j},
\label{eq:bcglobal}
\end{equation}
where $C$ is any possible choice of closed contour $C$ in the plane,
$L[C]$ is the winding number of the quantum phase variable 
$\theta(u,\phi)=\theta_0(u,\phi)$ (in the vortex gauge) associated to
that contour, and $\Phi[C]$ is the magnetic flux---normalized to the quantum 
of flux $\Phi_0$---through the surface $S$ in the plane bounded by $C$,
\begin{equation}
\Phi[C]=\frac{1}{2\pi}\int_S du\,d\phi\, ub(u,\phi).
\end{equation}
In particular, for a circular contour centered on the origin $u=0$ of 
the plane and of radius $u_0$, we thus have the condition
\begin{equation}
\frac{1}{2\pi}\int_0^{u_0}du\int_{-\pi}^{+\pi}d\phi\,ub(u,\phi)=
\Phi[u_0]=L[u_0]+\frac{1}{2\pi}\int_{-\pi}^{+\pi}d\phi\,g(u_0,\phi),
\label{eq:fluxcircle}
\end{equation}
where $L[u_0]$ stands of course for the associated winding number
in $\theta(u_0,\phi)$.

In the case of the infinite plane, the local boundary conditions at infinity
which complete those in (\ref{eq:bcglobal}) and which are
required for finite energy configurations are such that $f(u,\phi)$ approaches
either one of the two values $f=\pm 1$, 
while the quantities $j(u,\phi)$ and $g(u,\phi)$ must both vanish, since
they directly define the components of the electromagnetic current
density $\vec{J}=-f^2\vec{j}$ in the plane.

In the case of a bounded domain in the plane, there may exist an applied
external magnetic field $\vec{b}_{\rm ext}$ solely transverse to the
plane with a component $b_{\rm ext}$. Based on Maxwell's equations in vacuum
outside the bounded domain, in the specific instance of such a geometry
for $\vec{b}(u,\phi)$ it then follows that the magnetic field 
$\vec{b}(u,\phi)$
must indeed retain a constant homogeneous value throughout space in that
region. Consequently, given the remaining boundary conditions relevant
to $f^2\vec{j}$ and $\vec{\partial}f$ discussed previously, it then
follows that in the case of the disk topology the required local boundary
conditions which complete those in (\ref{eq:bcglobal}) are
\begin{equation}
b(u_b,\phi)=b_{\rm ext}\ \ ,\ \ 
j(u_b,\phi)=0\ \ ,\ \ 
\left(\partial_uf\right)(u_b,\phi)=0.
\label{eq:bclocal}
\end{equation}

In the case of the annular topology, the same boundary conditions apply 
of course at the outer boundary at $u=u_b$. Similarly at the inner boundary
$u=u_a$, one must have
\begin{equation}
b(u_a,\phi)=b_a\ \ ,\ \ 
j(u_a,\phi)=0\ \ ,\ \ 
\left(\partial_uf\right)(u_a,\phi)=0,
\label{eq:bcannulus}
\end{equation}
where $b_a$ represents the value of the constant magnetic field (transverse
component) within the hole of the annulus. The value for $b_a$ is to be 
determined from the total magnetic flux $\Phi[u_a]$ through that hole which 
is to be expressed in terms of the angular integral of $g(u_a,\phi)$ and the
winding number $L[u_a]$ in the quantum phase $\theta(u_a,\phi)$ of the order 
parameter $\psi(u_a,\phi)$ on the inner boundary, as given in 
(\ref{eq:fluxcircle}) with $u_0=u_a$. 
Note also that the fact that $b(u,\phi)$ must be
$\phi$-independent on the boundaries is consistent with the determination
of $b(u,\phi)$ inside the superconductor in terms of $j(u,\phi)$ and
$g(u,\phi)$ and with the boundary conditions on $j(u,\phi)$
(see (\ref{eq:MAX2})).

Finally let us consider the expression for the free energy of the
system. Due to the invariance of the considered configurations under
translations transverse to the plane, the three dimensional integral
in (\ref{eq:freeen}) is in fact infinite. We should thus rather consider
the expression for the free energy per unit of length in the
transverse direction, while we may also absorb the overall normalization
factor in (\ref{eq:freeen}) in that choice, thus leading in those units
to the expression
\begin{equation}
{\cal E}=\int_0^\infty du\int_{-\pi}^{\pi}d\phi\,
u\left\{\left[b-b_{\rm ext}\right]^2+
\left(\partial_uf\right)^2+\frac{1}{u^2}\left(\partial_\phi f\right)^2+
f^2\left(j^2+\frac{1}{u^2}g^2\right)+
\frac{1}{2}\kappa^2\left(1-f^2\right)^2\right\}.
\label{eq:free2}
\end{equation}
Note that the constant term $(-\kappa^2/2)$ has not been added to the
integrand of this expression, keeping in mind the possible infinite
planar topology in some applications (for the finite disk and annular
topologies, this term will be included later on). Note also that even though
the radial integration extends throughout the infinite plane, it is only
within the domain of the superconductor that $f$, $j$ and $g$ do not vanish,
while in the outer region of the supercondutor (if $u_b$ is finite)
we always have $b=b_{\rm ext}$, so that in fact there is no contribution
to the free energy for $u>u_b$ in the cases of the disk and annular
topologies (there is one for $u<u_a$ in the annulus case).
Actually, when evaluated for a specific solution to the differential
equations (\ref{eq:LG1}) and (\ref{eq:MAX2}), 
the expression for the free energy as given in (\ref{eq:free2})
also reduces to 
\begin{equation}
{\cal E}=\int_0^\infty du\int_{-\pi}^\pi d\phi\ u
\left\{\left[b-b_{\rm ext}\right]^2-\frac{1}{2}\kappa^2f^4+\frac{1}{2}\kappa^2
\right\}.
\end{equation}

The above set of equations and boundary conditions thus defines the
problem to be solved. The only information missing is that which provides
the topological structure of the vortex and domain wall configuration
which is being considered. This structure is cha\-rac\-te\-ri\-zed by the
set of winding numbers $L[C]$ which appear in the global boundary
conditions (\ref{eq:bcglobal}). The gauge dependent details of that
topological structure are specified through the choice of function
$\theta_0(u,\phi)$, which in the vortex gauge (\ref{eq:gaugefixing}) 
may be expressed as in (\ref{eq:theta0}) as explained previously,
and in terms of which the quantities $\theta(u,\phi)=\theta_0(u,\phi)$
and $\vec{a}(u,\phi)=\vec{j}-\vec{\partial}\theta_0$ are then obtained.
Nevertheless, it is only the gauge invariant content of $\theta_0$,
through the winding numbers $L[C]$, which is involved in the resolution
of the remaining quantities $f(u,\phi)$, $j(u,\phi)$ and $g(u,\phi)$,
which are to be determined from the above differential equations and 
boundary conditions, while also keeping in mind the double sheeted covering
of the plane which is associated to the vortex and domain wall
configuration being considered and over which the function $f(u,\phi)$
is to be defined.

As a final point, consider the case of the infinite plane or the disk
with a vortex of winding number $L_0\ne 0$ placed exactly at its center $u=0$, 
and a circular contour of radius $u_0$ surrouding that vortex. In the limit
that the contour shrinks to the origin, since the magnetic flux $\Phi[u_0]$
must vanish---the magnetic field being finite at $u=0$---,
the azimuthal component of the current density $\vec{J}=-f^2\vec{j}$
at that point must be such that
\begin{equation}
\lim_{u_0\rightarrow 0}\frac{1}{2\pi}\int_{-\pi}^\pi d\phi\,g(u_0,\phi)=
-L_0,
\end{equation}
which necessarily requires
\begin{equation}
g(0,\phi)=-L_0.
\end{equation}
On the one hand, since $j_\phi(u,\phi)=g(u,\phi)/u$, this result shows that
the current $\vec{j}$ is necessarily ill-defined at the location of
vortices. But on the other hand, since the order parameter $\psi$ and
the function $f$ vanish at the location of the vortex, while the
azimuthal component of the electromagnetic current $\vec{J}$ must
vanish at the same point (this is a consequence of the
polar coordinate parametrization which is being used),
this result also shows that the singularity
in $\vec{j}=-\vec{J}/f^2$ is just mild enough to be screened by the zero 
in $\psi$ in such a way that the physical current $\vec{J}$ remains 
well-defined and finite throughout space as it should,
even at the location of vortices.

\subsection{BPS bounds}
\label{Subsect3.2}

The resolution of the LGH equations, even in the plane, requires
a numerical analysis, since no analytic solutions are known.
Nevertheless, it is possible to establish some general properties
for the possible solutions in some particular situations. Namely,
for specific values of the scalar self-coupling $\kappa$ or $\lambda_0$,
it is possible to establish BPS lower bounds\cite{Bomo,Prasad} on the energy of
vortex configurations, as recalled in the Introduction.

To this aim, let us consider the free energy expressed in (\ref{eq:free2})
in the case that no external magnetic field is applied,
$b_{\rm ext}=0$, and for the disk and infinite planar topologies\footnote{The
annulus case is left to the reader.}. Through integration by parts and
the identification of the geometric meaning of the induced surface
terms, one then finds that
\begin{equation}
\begin{array}{r c l}
{\cal E}&=&\eta\left[2\pi\sum_{k=1}^KL_k+
\oint_{\partial\Omega}d\vec{u}\cdot\left(1-f^2\right)\vec{j}\right]\\
 & & \\
& &+\int_\Omega d^2\vec{u}\left\{
\left[\left(\frac{1}{u}\partial_ug-\frac{1}{u}\partial_\phi j\right)
-\frac{1}{2}\eta\left(1-f^2\right)\right]^2+
\left[\partial_uf+\eta\frac{1}{u}fg\right]^2+
\left[\frac{1}{u}\partial_\phi f-\eta fj\right]^2\right\}\\
 & & \\
 & & +\frac{1}{2}\left(\kappa^2-\frac{1}{2}\right)
\int_{\Omega}d^2\vec{u}\left(1-f^2\right)^2,
\end{array}
\label{eq:BPSenergy}
\end{equation}
where $\eta=\pm 1$ is some choice of sign to be specified presently,
$\Omega$ stands for the planar domain in which $f\ne 0$, namely that
of the superconductor, and $\partial\Omega$ for the boundary of that domain.
Finally, $L=\sum_{k=1}^KL_k$ is the total winding number of the
vortex configuration lying within the domain $\Omega$.

The remarkable feature of this relation is that, with the exception of the
surface term at $\partial\Omega$ and the
very last term which depends on the self-coupling $\kappa$, this expression 
gives the energy of any solution in terms of a sum of integrated 
positive quantities which only involve first-order variations---as does 
also the expression in (\ref{eq:free2})---but
in such a way that the topological content of the configuration is also
made explicit through the total winding number $L$. Furthermore,
when attempting to minimize the value of the energy by setting to zero
the positive integrated quantities, thus leading to the following
first-order differential equations,
\begin{equation}
b=\frac{1}{u}\partial_ug-\frac{1}{u}\partial_\phi j=\frac{1}{2}\eta
\left(1-f^2\right)\ ,\
\partial_uf=-\eta\frac{1}{u}fg\ ,\
\partial_\phi f=\eta ufj\ ,
\label{eq:firstorder}
\end{equation}
it is immediate to verify that these equations then also imply
the second-order equations in (\ref{eq:LG1}) and (\ref{eq:MAX2}),
but only if the coupling takes the specific value
$\kappa=\kappa_c$ with $\kappa_c=1/\sqrt{2}$. In other words,
when the scalar self-coupling takes the critical value $\kappa=\kappa_c$,
the system of first-order equations (\ref{eq:firstorder}) integrates
the second-order LGH equations of motion (the issue of boundary
conditions will be addressed shortly). Note also that it is for the
same critical coupling $\kappa_c$ that the very last term in
(\ref{eq:BPSenergy}) does not contribute to the energy of such configurations.

In order to fix the choice of sign for $\eta=\pm 1$, let us
note that the lowest possible value taken by ${\cal E}$ as defined 
in (\ref{eq:free2}) is zero, corresponding to the trivial solution
$b(u,\phi)=b_{\rm ext}$ (here $b_{\rm ext}=0$), $f(u,\phi)=\pm 1$,
$j(u,\phi)=0$ and $g(u,\phi)=0$ in the absence of any vortex in the plane.
Consequenly, all other solutions must necessarily possess a strictly
positive energy value, which dictates the following choice of sign
\begin{equation}
\eta={\rm sign}\left(\sum_{k=1}^KL_k\right)={\rm sign}\,L.
\end{equation}
Hence, given this choice for $\eta$ and the specific value
$\kappa=\kappa_c$, the energy of any solution to the first-order
equations (\ref{eq:firstorder})---which then also obey the
second-order ones---is given by
\begin{equation}
{\cal E}=2\pi\left|\sum_{k=1}^KL_k\right|+
\left({\rm sign}\,L\right)\ \oint_{\partial\Omega}d\vec{u}\cdot
\left(1-f^2\right)\vec{j}.
\end{equation}
However, it is only in the case of the infinite plane that the surface
term on the boundary $\partial\Omega$
does not contribute, since the boundary conditions
are then such that both $(1-f^2)$ and $\vec{j}$ vanish at infinity.
Furthermore, it is clear that these boundary conditions are also consistent
with the first-order equations (\ref{eq:firstorder}). Hence finally,
one concludes that all solutions to the first-order equations
(\ref{eq:firstorder}), which also obey the second-order LGH equations
and the appropriate boundary conditions
in the infinite plane when $\kappa=\kappa_c$,
saturate the BPS value which determines
their energy or mass in terms of a topological invariant, namely the total 
winding number of that configuration,
\begin{equation}
{\cal E}=2\pi\left|\sum_{k=1}^KL_k\right|=2\pi|L|.
\label{eq:BPS}
\end{equation}
Conversely, it has been shown\cite{Taubes} that under both the assumptions of
an everywhere positive function $f$ and possessing only a discrete set of
zeroes of integer positive degree, all the solutions to the
second-order LGH equations and boundary conditions
in the infinite plane with $\kappa=\kappa_c$ 
also solve the first-order equations (\ref{eq:firstorder}) 
with $\eta={\rm sign}\,L$. In other words, for the critical scalar 
self-coupling $\kappa=\kappa_c$, any solution to the LGH equations in the
infinite plane with 
isolated integer vortices only and of fixed total winding number $L$
saturates the BPS value above, thus also showing\cite{EWeinberg} 
that such vortices do not 
possess an interaction energy which would depend on their relative positions.

However, when relaxing these restrictions on the function $f$,
and thereby accounting for the possibility of half-integer vortices
and bound domain walls in the infinite plane, 
one must conclude that such configurations cannot
solve the first-order equations (\ref{eq:firstorder}) and thus cannot saturate
the BPS value (\ref{eq:BPS}). Indeed, the first of these first-order
equations would imply that on the surface of vanishing order parameter, 
$f=0$, within any domain wall, the magnetic field $b$
would retain a constant value of $\eta/2$ irrespective of the distance
between the vortices which are bound onto the edges of that domain wall
and irrespective of the winding numbers of these vortices.
Such a property seems very unlickely, not only because of this lack of
dependency of the magnetic field on the separation between the vortices at
the edges of domain walls and on their winding numbers, but also
because the magnetic energy density contribution to ${\cal E}$ would
then grow linearly with that distance, a fact which would clearly
be inconsistent with the saturation of the BPS value. 
On the other hand, we also know that the further a domain wall is streched 
(provided the vortices bound onto its two edges no longer overlap), 
the more the condensation energy stored into the domain wall contributes 
to ${\cal E}$ in an almost linear fashion as well, again
a property which would be incompatible with the saturation of the
BPS value (\ref{eq:BPS}). Hence, based on such physics arguments,
it must be concluded that even for the critical value $\kappa=\kappa_c$,
only isolated integer vortex solutions to the LGH equations in the infinite
plane both solve the first-order equations (\ref{eq:firstorder}) and saturate
the BPS value, while any other vortex configuration
including thus domain walls cannot share these properties. 
Nonetheless, such configurations with domain walls bound to vortices
in the infinite plane do obey the following BPS strict lower bound
\begin{equation}
{\cal E}>2\pi\left|\sum_{k=1}^KL_k\right|=2\pi|L|.
\label{eq:BPSstrict}
\end{equation}
Indeed considering again (\ref{eq:BPSenergy}), the surface term at infinity 
vanishes in the infinite plane while the first two-dimensional
volume integral is
strictly positive since such configurations cannot obey the first-order
equations (\ref{eq:firstorder}) (the last term vanishes for $\kappa=\kappa_c$).

Let us now turn to the disk of finite radius $u_b$.
In this case, one is led to the conclusion that the BPS lower bound 
${\cal E}\ge 2\pi|L|$
does not apply for solutions to the second-order LGH equations
with $\kappa=\kappa_c$, even for isolated integer vortices. 
The first-order equations
(\ref{eq:firstorder}) are in fact incompatible with the
required boundary conditions (\ref{eq:bclocal})\cite{Gov1}. Indeed, even when
$b_{\rm ext}=0$, these equations and boundary conditions together imply that
\begin{equation}
f(u_b,\phi)=\pm 1\ \ ,\ \ 
g(u_b,\phi)=0\ \ ,\ \ 
(\partial_\phi f)(u_b,\phi)=0.
\end{equation}
Clearly if $\sum_{k=1}^KL_k=L\ne 0$, such conditions can be met only if
the boundary of the disk is at infinity, $u_b\rightarrow\infty$,
thereby recovering the previous discussion in the infinite plane.
This conclusion on its own however, would still not exclude the
relevance of the BPS lower bound to solutions in the disk. In fact,
even had the first-order equations been
consistent with the boundary conditions, there cannot exist
a BPS lower bound on the
energy of any solution in a finite disk, including the case of
isolated integer vortices. Indeed, the surface
contribution $\oint_{\partial_\Omega}d\vec{u}\cdot(1-f^2)\vec{j}$
does not vanish in such a situation, and could {\sl a priori\/} carry
whatever sign depending on the details of the distribution of vortices
within the disk.

The situation with respect to possible BPS lower bounds having been
understood in the case of the critical coupling $\kappa=\kappa_c$, let
us now consider again the general expression (\ref{eq:BPSenergy})
relevant whatever the value for $\kappa$. Clearly, for couplings
larger than the critical one, $\kappa>\kappa_c$, and configurations
in the infinite plane thus leading to a vanishing surface term at infinity,
the energy always obeys the BPS strict lower bound in (\ref{eq:BPSstrict})
whatever the solution to the LGH equations. Indeed, on the one hand,
the very last contribution in (\ref{eq:BPSenergy}) proportional to 
$(\kappa^2-1/2)$ is then always strictly positive while, on the other hand,
the remaining integrated positive terms cannot vanish either since 
otherwise this would require the first-order equations (\ref{eq:firstorder})
(with $\eta={\rm sign}\,L$) to be obeyed by solutions to the LGH equations,
which is excluded when $\kappa\ne\kappa_c$.
In the case of the finite disk with $\kappa>\kappa_c$, again no
BPS lower bound may be given because of the non vanishing contribution of 
either sign from the surface term on the disk boundary. Values for the
energy ${\cal E}$ for configurations in the disk with $\kappa>\kappa_c$ could
{\sl a priori\/} be smaller as well as larger than the BPS value
$2\pi\left|\sum_{k=1}^KL_k\right|=2\pi|L|$ depending on the distribution
of vortices.

Finally, when $\kappa<\kappa_c$, the same conclusion as to the absence
of a BPS bound must be drawn, whether for vortex configurations in the
infinite plane or the disk, since in that case, the contribution of the
very last term in (\ref{eq:BPSenergy}) is always strictly negative.

\subsection{Singular BPS states}
\label{Subsect3.3}

There exists however, a specific situation, albeit a singular one,
for which all solutions to the LGH equations, including domains
walls, do in fact saturate the BPS lower bound (\ref{eq:BPS}).
As the above discussion has shown, one crucial property for saturating
the BPS bound is that the last integral in (\ref{eq:BPSenergy}) vanishes
identically\cite{Bomo,Prasad}. 
This may happen in either one of two ways, namely if 
$\kappa=\kappa_c$
as already discussed, or else if $f^2=1$ throughout the plane except
possibly on a subset of zero measure. Furthermore, if the latter
situation is realized, then the energy becomes independent of $\kappa$,
including even an infinite value for that coupling.

In the context of superconductivity, having chosen to normalize
distance measurements to the penetration length $\lambda$, the
limit $\kappa\rightarrow+\infty$ corresponds to a situation such that,
compared to $\lambda$, the coherence length $\xi$ becomes infinitely
small. Likewise, had we chosen to normalize length scales to $\xi$, the
same limit would correspond to a situation such that, compared to $\xi$,
the magnetic field penetration length becomes infinite. Hence, given
that the external field $b_{\rm ext}$ is assumed to vanish in our
present considerations, one should expect that in the limit 
$\kappa\rightarrow+\infty$ the magnetic field
$b(u,\phi)$ must vanish within the superconductor except possibly at
a few discrete points. As we shall see, this is indeed what happens.

In a particle physics context however, an infinite value for $\kappa$ would
likewise imply either a vanishing gauge boson mass $M_\gamma$ or else an
infinite Higgs mass $M_h$. In the first instance, the vacuum expectation
value $a$ would have to vanish while sending $\lambda_0$ to infinity in
order to keep $M_h$ non vanishing and finite, while in the second instance
the parameter $a$ is kept non vanishing and finite while $\lambda_0$ is also
sent to infinity. Thus in either case, this limit corresponds to an
infinitely strongly interacting Higgs sector, a regime which is
incompatible with perturbative unitarity in the quantized field theory.
Nevertheless, in the exploration of classical solutions to the
equations of motion of the abelian U(1) Higgs model, the limit
$\kappa\rightarrow+\infty$ may still provide some useful insight.

With this motivation in mind, let us consider then the second-order
LGH equations in (\ref{eq:LG1}) and (\ref{eq:MAX2}) subjected to the
relevant local and global boundary conditions. In particular,
since only the LGH equation (\ref{eq:LG1}) involves the parameter $\kappa$,
the limit $\kappa\rightarrow+\infty$ may be taken in a consistent manner
by first dividing that equation by $\kappa^2$. Consequently, the only
allowed solutions must be such that either $f=0$ or $f=\pm 1$ at
every point of the plane, precisely allowing for all types
of vortex and domain wall configurations defined over a double covering
of the plane or a finite domain of it. 
Considering then the remaining
equations for $b(u,\phi)$ as well as the global boundary conditions
(\ref{eq:bcglobal}), it then follows that given a vortex distribution
of winding numbers $L_k$ and positions $\vec{u}_k$ ($k=1,2,\cdots,K$)
in the plane or a finite domain of it, the solution to the LGH
equations in the limit $\kappa\rightarrow+\infty$ is given by
\begin{equation}
b\left(\vec{u}\right)=2\pi\sum_{k=1}^KL_k\delta^{(2)}
\left(\vec{u}-\vec{u}_k\right)\ \ ,\ \ 
\vec{j}\left(\vec{u}\right)=-\vec{\partial}\times\vec{b}\left(\vec{u}\right)
\ \ ,\ \ 
\psi\left(\vec{u}\right)=f\left(\vec{u}\right)e^{i\theta_0(\vec{u})},
\end{equation}
where the function $f(\vec{u})$ thus takes the value $f=0$ at the position
of vortices and domain walls, and otherwise either one of the values
$f=\pm 1$ according to the double sheeted covering of the plane which is
associated to that bound vortex and domain wall configuration.

Obviously, such solutions are singular precisely at the positions of vortices
and domain walls, with the magnetic field vanishing everywhere except at
the vortices where it diverges, and the function $f$ constant 
everywhere---on the double sheeted covering of the plane---except 
at its branch points and on its branch cuts where it vanishes
discontinuously---thus displaying the singularities of the double
sheeted covering of the plane. Furthermore, the current $\vec{j}$ and 
thus also the
electroma\-gne\-tic current $\vec{J}=-f^2\vec{j}$, vanishes everywhere, except
again at the vortices where it is singular. The physical picture however,
is clear. In the limit $\kappa\rightarrow+\infty$, the coherence length
$\xi$ being zero or the penetration length $\lambda$ being infinite,
the thickness of domain walls has shrunck to nothing while likewise
vortices have been squeezed 
into single curves without any transverse extension but nevertheless
each carrying the required finite value of magnetic flux, by having
``sucked in" all the currents responsible for that magnetic flux. In other
words, in the limit $\kappa\rightarrow+\infty$, the system of vortices
and bound domain walls has reduced to a collection, on the one hand,
of membranes of zero thickness which carry no magnetic flux, 
are bound at their edges to Aharonov-Bohm flux lines of integer or 
half-integer flux or extend to the boundary in the case of a finite 
domain in the plane, and which, on the other hand, may possibly also 
appear in combination with some isolated Aharonov-Bohm flux lines of 
integer flux.
Hence once again, such configurations are very much
reminiscent of D-branes\cite{Polchinski} and strings in M-theory, which 
are precisely such objects without any transverse extension.

Let us now consider the expression (\ref{eq:BPSenergy}) for the energy
of these configurations. Since the magnetic field $b$, the current
$\vec{j}$ and the Higgs potential $(1-|\psi|^2)^2$ all vanish
identically except on a subset of the plane of zero measure, all such
configurations of Aharonov-Bohm flux lines and membranes of fixed
total winding number $L=\sum_{k=1}^KL_k$ indeed saturate the
BPS lower bound, ${\cal E}=2\pi|L|$, thus showing that all
such solutions are indeed BPS states, albeit singular ones. In particular,
this result also implies that such Aharonov-Bohm flux lines and
membranes have no interactions with one another, whatever their
relative positions. From the physics point of view again, this fact
is rather obvious. The order parameter $\psi$ reaches 
its constant vacuum expectation value $|\psi|=1$ and
the magnetic field vanishes everywhere in space except at the positions
of flux lines and membranes (the latter applies to $|\psi|$ only), 
so that indeed both the magnetic and
condensation energy densities vanish identically irrespective of the
distribution of flux lines and membranes in space. Note that these
conclusions as to the BPS character of the solutions to the LGH equations
in the limit $\kappa\rightarrow+\infty$ remain valid whether in the
infinite plane or a disk of finite radius $u_b$. The only restriction in
the case of the infinite plane is that the total winding number 
$L=\sum_{k=1}^KL_k$ must then be an integer, without any membrane 
extending up to infinity.

\section{Half-Integer Vortex Solutions}
\label{Sect4}

Having argued that the basic entities from which to build general
solutions to the LGH equations are $1/2$-vortices, $1/2$-domain walls
and annular current flows, in this section we consider the situation
of a single vortex at the center of either a disk or an annulus
of finite radius. Even though no exact analytical solutions exist,
we shall try nevertheless to gain some insight into the nature of
such solutions by using some approximations, and then turn to the
results of a modest first attempt at a numerical analysis of half-integer
vortices.

When solutions with half-integer winding number are being considered,
note that we shall only consider here the case of a single domain wall
bound on such a vortex, and further assume that this domain wall lies
along one of the radii of the disk or annulus. Indeed, even though
such configurations necessarily break the axial symmetry of the disk
or annulus, this breaking is kept to a minimum under such a restriction,
which then remains covariant under rotations around the center of the
disk or annulus which map such solutions into one another without
changing their energy. This restriction is also in keeping with the
fact that the natural tension which domain walls possess is such that
in their lowest energy configuration domain walls must be straight not
only in the direction transverse to the plane, but also within that plane.

\subsection{Annular vortices}
\label{Subsect4.1}

In order to argue for the possibility of annular vortices of integer
as well as half-integer winding number, we shall use the same approach
as that of Ref.\cite{Gov1} which uncovered the existence of such configurations
in the isolated integer case by considering the emergence of these solutions 
from the trivial solution 
at the normal-superconducting phase transition. Namely, we shall 
restrict to the disk topology with a single vortex of winding number $L$
at its center, in the absence of any external magnetic field. Clearly,
the solution to the LGH equations which describes the system at the
phase transition is then given by
\begin{equation}
g(u,\phi)=-L\ \ ,\ \ j(u,\phi)=0\ \ ,\ \ b(u,\phi)=0\ \ ,\ \ f(u,\phi)=0,
\label{eq:normal}
\end{equation}
so that the order parameter vanishes identically, $\psi(u,\phi)=0$, with
a vanishing energy for the state, ${\cal E}=0$ (for the remainder of
the paper, the subtraction constant $(-\kappa^2/2)$ is being included
in the definition of the free energy (\ref{eq:free2})), and a non 
vanishing winding number nonetheless.

This solution may also be obtained by considering a rescaling by a real
parameter $f_0$ of the order parameter, such that
\begin{equation}
\psi(u,\phi)=f_0\,\tilde{\psi}(u,\phi)\ \ ,\ \ 
f(u,\phi)=f_0\,\tilde{f}(u,\phi),
\end{equation}
and then taking the limit $f_0=0$ in the LGH equations and relevant
boundary conditions. Doing so, one finds that the
function $\tilde{f}(u,\phi)$ satisfies the linearized LGH equation,
\begin{equation}
\left[u\partial_uu\partial_u+\partial^2_\phi\right]\tilde{f}=
\left[(uj)^2+g^2\right]\tilde{f}-\kappa^2u^2\tilde{f}.
\end{equation}
Given the solution in (\ref{eq:normal}), as well as the requirement that
$\tilde{f}(u,\phi)$ should vanish at the origin when $L\ne 0$ since this 
is the position of the vortex where $f(u,\phi)$ must vanish, the linearized
LGH equation possesses a single solution. The case of an isolated vortex
of integer winding number has been discussed in Ref.\cite{Gov1}. 
Let us restrict here to that of a half-integer value for the winding number 
$L$ of a single vortex which is bound onto a single domain wall lying
along a radius of the disk (the generalization to
more domain walls and integer winding numbers is immediate, 
as the reader will realize).

Under these specific assumptions, and the fact that
the equation for $\tilde{f}(u,\phi)$ in the limit $f_0=0$ is 
linear, it is natural to consider the following
separation of variables,
\begin{equation}
\tilde{f}(u,\phi)=\tilde{f}(u)\sin(\phi/2),
\end{equation}
which clearly displays a cut at $\phi=\pm\pi$,
having assumed the domain wall to be at $\phi=0$ where the branch cut
for $\tilde{f}(u,\phi)$
in the double covering of the disk thus lies. The solution for
$\tilde{f}(u)$ is then simply a Bessel function of the first kind
with index $\alpha=\sqrt{L^2+1/4}$,
\begin{equation}
\tilde{f}(u)=\left(\frac{2}{\kappa}\right)^\alpha\,\Gamma(1+\alpha)
J_\alpha(\kappa u),
\end{equation}
where the normalization is chosen such that the lowest order term
in a series expansion in $u$ is $\tilde{f}(u)\simeq u^{\alpha}$. 
As is well known, such a Bessel function has an almost periodic
oscillatory behaviour with an amplitude
which asymptotically decreases as $1/\sqrt{u}$ at large radii.

Consequently, as soon as the parameter $f_0$ is turned on slightly
away from zero, this oscillatory pattern hidden in the solution
$f(u,\phi)=0$ will emerge from the vanishing condensate, and lead
to a pattern of concentric annular current flows of approximately
constant width $\pi/\kappa$
(in units of $\lambda$) surrounding the half-integer
vortex of winding number $L$ which is bound onto the edge of a single
domain wall (only the index $\alpha$ of the Bessel function as well as the
periodicity of the trigonometric function in $\phi$ change when
more domain walls are involved). Of course, the boundary conditions 
at $u=u_b$ are then no longer satisfied, so that both the parameter $f_0$
as well as the remaining functions $j$ and $g$ must then be adjusted
in order to obtain a solution of negative energy to the LGH equations,
thereby also pushing outwards the pattern of concentric annular current
flows without essentially changing their width\cite{Gov1}.

Nevertheless, the existence of the Bessel function solution to the
linearized LGH equation at the phase transition
provides a very strong argument for the existence in finite domains
of annular vortex solutions of arbitrary half-integer winding
number bound onto the edges of an arbitrary number of domain
walls. The existence of such solutions without domain walls
has indeed been established
along such lines for integer vortices in Ref.\cite{Gov1}. There is no
reason to doubt that the same conclusion should not also hold for
integer and half-integer vortices bound onto domain walls.

Even though it is impossible to obtain an exact solution to the equations
when turning on the parameter $f_0$, even in the form of
a power series expansion, 
it is interesting to consider the first-order corrections
that this implies for the quantities $j$, $g$ and $b$, without including
the corrections implied for $f(u,\phi)$ itself. To lowest order in $u$, 
and still assuming only a single domain wall, one finds, when also imposing
the current conservation equation,
\begin{equation}
\begin{array}{r c l}
uj(u,\phi)&=&\frac{L}{2\alpha}\frac{\cos(\phi/2)}{\sin(\phi/2)},\\
g(u,\phi)&=&-L+\frac{1}{2}b_0u^2-\frac{L}{4\alpha(\alpha+1)}f^2_0
u^{2(\alpha+1)}\sin^2(\phi/2)-\frac{L}{2\sin^2(\phi/2)}\ln(u/u_0),\\
b(u,\phi)&=&b_0-\frac{L}{2\alpha}f^2_0u^{2\alpha}\sin^2(\phi/2),\\
f(u,\phi)&=&f_0u^{\alpha}\sin(\phi/2),
\end{array}
\end{equation}
where $b_0$ stands for the value of the magnetic field at the vortex $u=0$
and $u_0$ for an arbitrary integration constant.
These expressions are at least indicative of the singularities in
$j(u,\phi)$ and $g(u,\phi)$ not only at the position of the
vortex---a fact which was already demonstrated previously---but also
on the surface of vanishing order parameter at $\phi=0$, namely on
the branch cut in the function $f(u,\phi)$, but in such a way that the
physical electromagnetic current $\vec{J}=-f^2\vec{j}$ remains well
defined and regular on this branch cut nonetheless, including the
quantum tunnel effect through the domain wall in the proximity of the vortex
as was described in the Introduction. These properties,
announced previously, are thus made explicit in this approximate
representation of a single vortex in the disk bound onto a single domain wall.

\subsection{The thin annulus limit}
\label{Subsect4.2}

Another instance for which it is possible to gain insight into the
nature of half-integer vortex solutions in a particular limit,
is that when the width of the annular topology vanishes, the
outer radius being kept fixed, namely the limit $u_a\rightarrow u_b$.
Let us assume that a single winding number $L$ is trapped in the
center of the annulus of width $\Delta u=u_b-u_a$ and outer radius $u_b$.
Consequently, the value $b_a$ of the magnetic field inside the hole
must be such that
\begin{equation}
\frac{1}{2\pi}\int_{-\pi}^{\pi}d\phi\,g(u_a,\phi)=\frac{1}{2}u^2_ab_a-L.
\end{equation}

In the limit that the annulus becomes infinitely thin, $u_a\rightarrow u_b$,
the magnetic field both within the superconductor as well as inside
the hole takes the value $b_{\rm ext}$ of the applied field. Moreover
in that limit we shall also take the approximation that $j(u,\phi)=0$,
the annulus having no radial extent anymore, even though such a
restriction may prove to be inconsistent with the current conservation 
condition.
In this limit, not only does the above global boundary condition imply 
a specific restriction on the function $g(u_b,\phi)$ in terms of $L$
and $b_{\rm ext}$, but also the free energy (\ref{eq:free2}) then
reduces to the expression,
\begin{equation}
{\cal E}\simeq\frac{\Delta u}{u_b}\int_{-\pi}^\pi d\phi
\left\{\left(\partial_\phi f\right)^2+f^2g^2+\frac{1}{2}\kappa^2u^2_bf^4-
\kappa^2u^2_bf^2\right\}.
\end{equation}

In order to address the problem of minimizing the free energy and thus
solve the LGH equations in the thin annulus limit, let first consider
the case of an integer winding number $L$ without any domain wall lying
along some radius within the annulus. Because of the axial symmetry of such
a configuration, no $\phi$-dependency arises for the functions
$b$, $g$ and $f$, and one is left only with the following relations
(with all these quantities evaluated at $u=u_b$, of course)
\begin{equation}
g=\frac{1}{2}u^2_bb_{\rm ext}-L\ \ ,\ \ 
{\cal E}\simeq 2\pi\frac{\Delta u}{u_b}
\left[f^2g^2+\frac{1}{2}\kappa^2u^2_bf^4-\kappa^2u^2_bf^2\right].
\end{equation}
It is immediate to show that the free energy is minimized for
the following values
\begin{equation}
f^2=1-\frac{1}{\kappa^2u^2_b}g\ \ ,\ \ 
{\cal E}\simeq -\frac{\pi\Delta u}{u_b}(\kappa^2u^2_b)
\left[1-\frac{1}{\kappa^2u^2_b}
\left(L-\frac{1}{2}u^2_bb_{\rm ext}\right)^2\right]^2.
\end{equation}
In turns out that this quartic dependency of ${\cal E}$
on the applied field $b_{\rm ext}$
provides a rather good approximation to exact numerical solutions for
annuli of finite width $\Delta u$ small compared to $u_b$, 
since it presents a turn over point precisely
at the values of $b_{\rm ext}$ where this expression vanishes. This property
is very close to the behaviour of the exact solutions, while it is
not reproduced by the quadratic dependency which is usually
discussed in textbooks\cite{Tinkham,Waldram}. Whatever the relevance of this latter remark,
the important point is that when considered for all possible integer
values $L$, the graphs for the free energy ${\cal E}$ as a function of
$b_{\rm ext}$ all cross one another in succession before their turn over
points at ${\cal E}=0$ provided that $\kappa u_b>1/2$ and for values
such that
\begin{equation}
\frac{b}{\xi}=\kappa u_b>\frac{1}{2}\ :\ \ \
\frac{1}{2}u^2_bb_{\rm ext}=k+\frac{1}{2}\ \ ,\ \ 
{\cal E}(k)\simeq -\frac{\pi\Delta u}{u_b}(\kappa^2u^2_b)
\left[1-\frac{1}{4\kappa^2u^2_b}\right],
\end{equation}
$k$ being an arbitrary positive, negative or zero integer.
This, of course, is the experimentally observed behaviour as well, 
namely the Little-Parks effect\cite{Tinkham,Waldram}, 
the crossing points occurring for a magnetic 
flux through the annulus equal to a half-integer multiple of the quantum of 
flux $\Phi_0$ (note also the lower bound on the annulus radius,
$b>\xi/2$). Furthermore, the absolute minimimum of the free
energy is reached for
\begin{equation}
\frac{1}{2}u^2_bb_{\rm ext}=k\ \ ,\ \ 
{\cal E}_{\rm min}\simeq -\frac{\pi\Delta u}{u_b}(\kappa^2u^2_b),
\end{equation}
while the absolute maxima are reached for
\begin{equation}
\frac{1}{2}u^2_bb_{\rm ext}=k\pm\kappa u_b\ \ ,\ \ 
{\cal E}\simeq 0.
\end{equation}

Let us now consider a similar analysis in the case of a half-integer
winding number $L$, which should thus also include 
some domain wall structure inside
the thin annulus. The axial symmetry of the equations then being broken,
all quantities $b(\phi)$, $g(\phi)$ and $f(\phi)$ become $\phi$-dependent.
Hence, the minimization of the free energy leads in this case to the
differential equation,
\begin{equation}
\frac{d^2}{d\phi^2}f=fg^2-\kappa^2u^2_b(1-f^2)f,
\end{equation}
while the function $g(\phi)$ is constrained by the condition
\begin{equation}
\frac{1}{2\pi}\int_{-\pi}^\pi d\phi g(\phi)=\frac{1}{2}u^2_bb_{\rm ext}-L.
\end{equation}
Finally, let us also include the current conservation equation, which
in the present limit reduces to
\begin{equation}
\partial_\phi(f^2g)=0.
\end{equation}
Hence, we must have
\begin{equation}
g(\phi)=\frac{C}{f^2(\phi)}\ \ ,\ \ 
\frac{d^2}{d\phi^2}f=\frac{C^2}{f^3}-\kappa^2u^2_b(1-f^2)f,
\label{eq:thinannulus}
\end{equation}
with $C$ some integration constant.

First in the case that this constant vanishes, $C=0$, we necessarily
have $g(\phi)=0$, which thus implies that such a situation is
obtained only for values of the applied field such that
\begin{equation}
\frac{1}{2}u^2_bb_{\rm ext}=k+\frac{1}{2}=L,
\end{equation}
$k$ being an arbitrary integer. These are precisely the values
for which the energy of the thin annulus is degenerate for two 
consecutive integer winding numbers (provided $\kappa u_b>1/2$). 
Furthermore, the solution for the order parameter is then
\begin{equation}
f(\phi)=\pm\tanh\left(\frac{\kappa u_b}{\sqrt{2}}\phi\right),
\end{equation}
leading to a value for the free energy of this configuration given by
\begin{equation}
{\cal E}\simeq -\frac{\pi\Delta u}{u_b}(\kappa^2u^2_b)
\left\{1-\frac{2\sqrt{2}}{\pi\kappa u_b}
\tanh\left(\frac{\pi\kappa u_b}{\sqrt{2}}\right)
\left[1-\frac{1}{3}\tanh^2\left(\frac{\pi\kappa u_b}{\sqrt{2}}\right)
\right]\right\}.
\end{equation}
This value always lies above that ${\cal E}(k)$
for the integer winding number case at the
same magnetic field values, whatever the value for $\kappa u_b>1/2$.

The solution found for $f(\phi)$ does indeed display a cut at
$\phi=\pm\pi$, in accordance with the half-integer winding number,
and vanishes at $\phi=0$, indicating the presence of a single domain wall
in the thin annulus. However, the condensate $|\psi(\phi)|^2=f^2(\phi)$
presents a discontinuity in its angular derivative at the cut at
$\phi=\pm\pi$, since we have
\begin{equation}
f^2(\phi)=\tanh^2\left(\frac{\kappa u_b}{\sqrt{2}}\phi\right)\ \ ,\ \ 
\frac{d}{d\phi}f^2(\phi)=\sqrt{2}\kappa u_b
\tanh\left(\frac{\kappa u_b}{\sqrt{2}}\phi\right)
\left[1-\tanh^2\left(\frac{\kappa u_b}{\sqrt{2}}\phi\right)\right].
\end{equation}
Such a discontinuity is not physically acceptable, and is a consequence
of the thin annulus limit which we have taken. Nevertheless, this
approximate solution to the full LGH equations provides some interesting
insight into the properties of half-integer vortex solutions.

Let us now turn to the situation when the integration constant $C$
does not vanish. In fact, the differential
equation to be solved, (\ref{eq:thinannulus}),
is also that of a non relativistic particle of unit mass
moving on the real line $-\infty<f<+\infty$ and subjected to the
potential
\begin{equation}
V(f)=\frac{C^2}{2f^2}+\frac{1}{2}\kappa^2u^2_b\left(f^2-\frac{1}{2}f^4\right),
\end{equation}
whose conserved energy $K_f$ is thus
\begin{equation}
\left(\frac{d}{d\phi}f\right)^2+V(f)=K_f.
\end{equation}
Hence, there cannot exist a solution for $f(\phi)$ passing
through the value $f=0$ for any finite value of the integration
constant $K_f$, since the potential $V(f)$ diverges at that point.
An infinite value for $K_f$ would rather be required, but this in turn
is physically unacceptable since in such a case the
angular variation $df/d\phi$ of the order parameter
would be infinite at all points except possibly at those where $f=0$,
thus also leading to an infinite energy.
Hence we must conclude that no solution to the above equations
may be found when $C\ne 0$.

The physical interpretation of these results is as follows. Having
taken the thin annulus limit $u_a\rightarrow u_b$ with the restriction
that $j(\phi)=0$, this is consistent with the current conservation
equation only for integer winding numbers without domain walls, since the axial
symmetry of such solutions even for an annulus of finite width is
such that $j(u,\phi)=0$ in any case. For half-integer winding numbers
or integer vortices bound to domain walls however, 
this limit is such that all states are squeezed out from the
annulus, by acquiring an infinite energy. It is only when the applied field
$b_{\rm ext}$ takes the distinguished values at which two consecutive
integer winding states without domain walls become degenerate (which requires
$\kappa u_b>1/2$), that
the states of half-integer winding number and with a single domain wall 
across the annulus retain
a finite energy, but yet develop a discontinuity in the angular
variation of the order parameter.

\subsection{Half-integer vortices in the disk}
\label{Subsect4.3}

Let us now finally turn to some exact solutions to the LGH equations 
(\ref{eq:LG1})
and (\ref{eq:MAX2}) in the disk, albeit numerical ones. The numerical
resolution of these differential equations presents some genuine challenges.
Even in the case of axially symmetric configurations, namely a single
isolated giant vortex of integer winding number at the center of the
disk, and thus in the absence of any domain wall, the direct numerical
integration of the then ordinary differential equations for $f(u)$, $g(u)$
and $b(u)$---$j(u)=0$ in that case---are quite delicate. These
equations then read
\begin{equation}
u\frac{d}{du}b(u)=f^2(u)g(u)\ \ ,\ \ 
u\frac{d}{du}\left[u\frac{d}{du}\right]f(u)=g^2(u)f(u)-\kappa^2u^2
\left(1-f^2(u)\right)f(u),
\end{equation}
with of course $ub(u)=dg(u)/du$ and the boundary conditions
\begin{equation}
\begin{array}{r c l}
g(0)=-L\ &;&\ \left(\frac{d}{du}f\right)(0)=0\ {\rm if}\ L=0\ ,\
f(0)=0\ {\rm if}\ L\ne 0,\\
b(u_b)=b_{\rm ext}\ &;&\ \left(\frac{d}{du}f\right)(u_b)=0\ \ ,
\end{array}
\end{equation}
$L$ being the integer winding number.
The main reason for difficulties in this case is, on the one
hand, that the equations are non linear---the cubic term in $f^3$ in the
LGH equation makes the stability of any numerical integration problematic---,
and on the other hand, that the above local boundary conditions are
in fact delocalized at the two boundaries of the disk, namely at
$u=0$ and at $u=u_b$. Hence, whether integrating inwards or outwards,
half of the initial values must be adjusted in order to satisfy the
required boundary conditions at the other boundary. Nevertheless,
it is still possible to manage such difficulties, and even obtain
the annular vortex solutions although the precision requirements
on the initial values then become critical\cite{Gov1}.

In the case of configurations which are no longer axially symmetric,
namely those involving either isolated integer vortices which are no longer
at the center of the disk, or domain walls, or both, these
difficulties are much greater, since the non linear equations are then
partial differential ones in two variables for a larger number of
quantities to be integrated, and with not only
delocalized local boundary conditions but also the global boundary
conditions corresponding to the integrated London equations for 
all contours in the plane. Furthermore, as we have seen, the
current $\vec{j}(u,\phi)$ is then also singular at the positions
of vortices and inside the domain walls. These singularities at the
location of a vortex which is at the center of the disk can be handled
as above in the axially symmetric case, since we then have $j(0,\phi)=0$ 
and $g(0,\phi)=-L_0$. Unfortunately, such a simplification does not apply
to vortices away from the origin $u=0$ nor to domain walls.

The alternative to integrating numerically the partial differential
equations (\ref{eq:LG1}) and (\ref{eq:MAX2}) subjected to the local
and global boundary conditions, is to consider rather the numerical
minimization of the free energy (\ref{eq:free2}) of the system
using a steepest descent method of one type or another. Indeed,
since the free energy coincides, up to a sign, with the action in the
case of time independent configurations, the minimization of the
energy is equivalent to solving the equations of motion. Most of the
difficulties mentioned above may then be circumvented, by an appropriate
choice of lattice discretization and of the positioning of the vortices and
domain walls off the lattice sites. Such a procedure however, requires
a lot of computational power, since all variables to be determined---in the
present case the quantities $j$, $g$ and $f$---are to be varied at each
lattice site in order to follow the path of steepest descent in the
free energy. This is, in a few words, the general method which 
has been applied in order to obtain the results to be presented here and in 
section \ref{Subsect5.2}. Further details will be provided elsewhere.

At this stage, simulations of only modest computer time have been run,
leading to results which lack great precision. The only results to be
described in this section apply for the following choice of parameters
\begin{equation}
u_b=3\ \ \ ,\ \ \ \kappa=1.
\end{equation}
The lattice discretization uses the polar coordinate parametrization
with 10 intervals in the radial variable $0\le u\le u_b$ and 10
intervals in the angular range $0\le\phi\le\pi$. Indeed, the only
configurations considered involve either a single integer vortex
with $L=0,1,2$ without a domain wall at the center of the disk, or else
a single half-integer vortex with $L=1/2,3/2$ also at the center
of the disk and then bound onto only one domain wall. Furthermore,
for reasons already given previously, we assume this domain wall to be
straight and to extent all the way to the boundary of the disk by lying
along one of its radii, which we take to be the one at $\phi=0$.
Consequently, all these configurations possess specific symmetry
properties under $\phi\rightarrow -\phi$, which may be taken
advantage of to reduce by half the calculation labor.

The rather coarse grained discretization used may imply a rather
poor precision. Since the same procedure is applied to all
configurations, one may reasonably hope that whatever the numerical
uncertainties, they would all contribute in the
same direction so that at least the general properties, if not the
precise numerical values, are to be trusted. Indeed, at least in
the case of the configurations with integer winding number which are
axially symmetric, the angular integration becomes trivial and a far
more precise resolution, either by minimizing the free energy or by
integrating the ordinary differential equations of motion using a
4th order Runge-Kutta method, becomes possible. The comparison between
the different approaches then shows that the steepest descent
method used in the half-disk leads to values for the free energy
which are precise to 1\% to 2\% but which all deviate from their actual
values in the same direction as expected. Hence, the same quality
should apply to the results for the configurations with $L=1/2$ and
$L=3/2$.

Fig.1 presents the results for the free energy of the disk in these
different vortex and domain wall configurations as a function of the
applied external magnetic field $b_{\rm ext}$. Of course, the well known
behaviour of the energy for the $L=0,1,2$ configurations is reproduced.
However, the curves for the solutions of winding numbers $L=1/2$
and $L=3/2$ are totally new, and are quite noteworthy. First, one
may notice that at the values of $b_{\rm ext}$ where the curves of integer
$L$ cross each other, the energy values for the half-integer winding numbers
always lie above those for the integer winding numbers, in agreement
with the result of the analysis in the thin annulus limit. Second,
the magnetic field dependency of the energy for half-integer winding
numbers seems to be less pronounced than for the integer ones, and thus
extends to larger values of $b_{\rm ext}$ before hitting the phase transition
line at ${\cal E}=0$. Nevertheless, these curves always lie above those
of integer winding numbers (including $L>2$ not
represented), so that the lowest energy states, at least
for the geometry considered, always correspond to some giant vortex
state without a domain wall. Third, the $L=1/2$ and $L=3/2$ curves lie
much closer to one another than do the integer winding number ones.
In particular, it is quite remarkable that at $b_{\rm ext}=0$, both
half-integer winding number curves lie below all the giant vortex ones
without domain wall for $L\ne 0$\footnote{In this respect, it would
be interesting to see where the curve for the $L=3/2$ with three domain
walls would lie.}.

Clearly, this is only one single example of such results in a very
specific case, and it would be extremely interesting to see how these
different features would change when varying the geometry of the disk
and the value for $\kappa$,
when moving the vortices around within the disk, when allowing
the domain walls to end on some other vortex inside the disk,
when including more domain walls, and so on. Obviously, all these
issues are of direct relevance to the detailed understanding of the
magnetization properties and dynamics of mesoscopic superconducting
disks and loops in varying electromagnetic fields.

Even though not displayed, the numerical solutions for the configurations
considered above also confirm the fact that there is indeed an
electromagnetic current flow which quantum tunnels through the
domain wall close to the vortex position for the $L=1/2,3/2$ winding
number cases. 
Furthermore, within the core of these vortices, the magnetic field
is of course not axially symmetric, but in fact it even presents a shape such
that its values first increase when moving out from the center of the
vortex before decreasing towards the disk boundary. This volcano-like shaped
surface also possesses a valley-like small depression in the direction of the
domain wall, along which the order parameter $|f|$ takes of course smaller
values and thus lets the magnetic field penetrate further in and decrease
less rapidly as well (see (\ref{eq:MAX2})).

In spite of the modesty of these first numerical results,
they do establish the existence of half-integer winding number solutions
to the LGH equations, bound to domain walls, and demonstrate the
potential these new configurations have to contribute in crucial
ways to the static and dynamic properties of any physical system which is
described, even if only in an effective way, by the abelian U(1) Higgs
model, and such instances of course are not confined to superconductors only.

\section{Bound $1/2$-Domain Walls}
\label{Sect5}

\subsection{The solitonic limit}
\label{Subsect.5.1}

Bound $1/2$-domain walls were argued to provide the basic
entities in terms of which any solution to the LGH equations
in the infinite plane could be viewed as a particular
combination of such states being stacked together. Before
considering in the next section the results of a modest attempt 
at a numerical resolution, here we shall present a specific
bound $1/2$-domain wall solution to the LGH equations (\ref{eq:LG1}) 
and (\ref{eq:MAX2}) in the infinite plane.

Even though this solution possesses an infinite energy, it may in fact
be viewed to correspond to a bound $1/2$-domain wall which is
infinitely streched in a specific direction so that the two $1/2$-vortices
bound onto its edges are at infinity. It is this infinite length of
the domain wall which explains its infinite energy of course. In spite
of this fact however, the solution is interesting in that it illustrates
some of the properties of bound $1/2$-domain walls in a particularly
simple limit, while also demonstrating that indeed the energy of
domain walls is linear in their length when the vortices at their
edges no longer overlap.

To describe this specific solution, it is best to consider a system
of cartesian coordinates $x$ and $y$ in the infinite plane
(normalized to $\lambda$), and to assume
that the domain wall is aligned, say, with the $x$ axis, so that the
configuration is also invariant under translations parallel to that axis. 
Hence, all variables must be independent of $x$. Furthermore,
given the physical interpretation, in which the $1/2$-vortices
are at infinity, one should expect that both the magnetic field
$b(y)$ as well as the current $\vec{j}(y)$, and thus also the
electromagnetic current density $\vec{J}(y)=-f^2(y)\vec{j}(y)$, 
vanish everywhere in the plane,
\begin{equation}
b(y)=0\ \ ,\ \ \vec{j}(y)=\vec{0}.
\end{equation}
These restrictions are clearly consistent with the current conservation
equation, while the LGH equation for the order parameter then reduces to
\begin{equation}
\frac{d^2}{dy^2}f(y)=-\kappa^2
\left(1-f^2(y)\right)f(y).
\end{equation}
This latter equation is of course well known, and possesses solitonic
solutions running from one vacuum $f=-1$ of the Higgs potential to the
other $f=+1$. In the present case, the solution we are interested in
is that corresponding to a single domain wall which separates two regions
of the plane in which each of these two expectations values are reached
at infinity.

Hence, the infinitely streched bound $1/2$-domain wall is described by
the one soliton or antisoliton configuration
\begin{equation}
f(y)=\pm\tanh\left(\frac{\kappa}{\sqrt{2}}y\right).
\end{equation}
Note how this solution takes values of opposite signs on both sides
of the $x$ axis, the choice of $\pm$ sign in this expression corresponding
to a choice of sheet in the double sheeted covering of the plane which is
defined by these one soliton and antisoliton configurations.
Similarly, a $N>1$ soliton (or antisoliton)
solution would correspond to $N$ such
infinitely streched bound $1/2$-domain walls all lying parallel
to one another in the infinite plane. However, depending on the odd or
even character of $N$, such states are not topologically stable and
would decay either to a single bound $1/2$-domain wall if $N$ is odd, or
else to the vacuum configuration $|f(y)|=1$ if $N$ is even.

Clearly, these configurations being translationally invariant along a
specific direction in the plane, possess an energy (\ref{eq:free2})
which is infinite. Nevertheless, let us consider their energy per unit
of length $\Delta x$ along their symmetry axis, chosen here to be the $x$ 
axis. One then finds for the one soliton 
configuration\footnote{The calculation
is also feasable if the plane is of finite extent in the $y$ direction,
$-y_0<y<y_0$, but the boundary conditions $(df/dy)(\pm y_0)=0$ are then
not satisfied.}
\begin{equation}
{\cal E}/\Delta x=\frac{4\sqrt{2}}{3}\kappa.
\end{equation}
Hence in this limit of an infinite separation of the two edges
of the bound $1/2$-domain wall, the energy does grow linearly with that 
distance. The amount of condensation energy---no magnetic energy density
is present in the configuration---stored in the domain wall which is of
constant thickness is proportional to its length. These are indeed the
expected properties of streched bound $1/2$-domain walls, as discussed
in the Introduction.

\subsection{The bound $1/2$-domain wall in the disk}
\label{Subsect5.2}

Let us now consider the numerical resolution of the LGH equations
for a bound $1/2$-domain wall in the disk topology, using the
minimization of the free energy through steepest descent. Only the
specific instance of a vanishing external magnetic field will be
considered, $b_{\rm ext}=0$. Clearly, such an analysis touches directly
onto the issue of the stability of the ANO vortex
of winding number $L=\pm 1$, so that we shall in fact consider
for different values of $\kappa$
the dependency of the energy (\ref{eq:free2}) of such states as
a function of the separation between the two $L=\pm 1/2$ vortices
bound onto the edges of such a domain wall.

In the Introduction, it was argued that there might exist a critical
value $\kappa_{1/2}$ marking the boundary between the situation where,
on the one hand, the bound $1/2$-domain wall would be stable for a specific
streched length (function of $\kappa$), and on the other hand, where it would
always collapse into the $L=\pm 1$ ANO vortex.
Furthermore, it was also suggested, based on the interplay between
repulsive and attractive magnetic and scalar forces and the situation
in this respect for isolated giant vortices, that this critical value
$\kappa_{1/2}$ could even coincide with the critical value $\kappa_1$
which governs a similar behaviour for integer vortices without
domain walls, and which also coincides with the critical value
$\kappa_c$ for which the latter states saturate the BPS bound.
Based on such considerations, one would thus expect that the
numerical resolution should possess some distinguishing feature when
$\kappa=\kappa_c=1/\sqrt{2}$. In particular, it could then be the case that,
for $\kappa<\kappa_c$ the energy of the bound $1/2$-domain wall
would display a strictly positive monotonous increase as a function of
its extension starting already from 
zero separation, while for $\kappa=\kappa_c$ this
same derivative would vanish, and finally for $\kappa>\kappa_c$, a small
dip in the energy would be observed close to the origin
corresponding to a split Abrikosov
vortex at equilibrium, before turning over into the
linear increase which is to be expected in any case for a sufficiently
large separation.

On the other hand, the existence of BPS lower bounds in the case of an infinite
plane may also be brought to bear on this issue, in fact then leading to 
the opposite expectations! Indeed, even when $\kappa=\kappa_c$, we have seen
that the only states which saturate the BPS bound are the isolated
integer giant vortex ones, whereas any of the other vortices which
involve at least one domain wall have an energy which is necessarily
always strictly larger. Hence, in the infinite plane, the $L=\pm 1$
ANO vortex must be stable when $\kappa=\kappa_c$.
For $\kappa>\kappa_c$ however, since the BPS lower bound can never
be saturated for whatever solution, the possibility that the energy
could present a minimal value at non zero separation, before turning
over into the linear increase, remains an open possibility. 
The same remark also applies when $\kappa<\kappa_c$,
the more so since no BPS lower bound applies in this case,
the term proportional to $(\kappa^2-1/2)$ in (\ref{eq:BPSenergy}) 
being then always strictly negative. In fact, from this point of view,
this is the situation which would seem to offer the best case
for an unstable ANO vortex since this term
is the more negative the more the domain wall is streched (of course,
the strictly positive two-dimensional volume integral in (\ref{eq:BPSenergy})
always ends up winning over for sufficient separation). These types
of arguments would thus indicate that the value $\kappa=\kappa_c$ could
indeed play a distinguished role, but rather in exactly the opposite
way to that which was argued for previously.

To make things even more delicate, let us also note that for the more 
realistic case of a finite planar domain, such as a disk or an annulus,
the BPS lower bounds no longer apply because of the contribution to the
energy of the surface term at the boundary, so that even for the value
$\kappa=\kappa_c$ the issue becomes uncertain. Furthermore, for finite
domains, when the domain wall is sufficiently streched so that 
at least one of 
its edges comes close to the boundary, the Bean-Livingston barrier\cite{Bean}
would also become involved in the energy balance of the system, leading to 
further structure in the dependency of the energy on the vortex separation.

Finally, let us also recall that in the singular limit 
$\kappa\rightarrow+\infty$,
all the solutions to the LGH equations become degenerate in energy
and saturate the BPS bound. Since there are no interactions acting between 
the then membranes and Aharonov-Bohm flux lines, the bound $1/2$-domain wall
is stable for whatever separation of its two edges. This remark
indicates that at least in this extreme regime, the ANO vortex
may indeed split into such a state.

Hence, one must conclude that given this lack of guidance for an
educated guess, only a detailed numerical analysis of the problem
can answer this all important issue of the stability of the
ANO vortex as a function of the coupling
$\kappa$. This section presents the results of a first attempt
towards a resolution, which, unfortunately, proves to be still too
modest to enable any definite conclusion. In fact, a full-fledged
dedicated computationally intensive numerical analysis is required
to resolve this important question, which we hope to be able
to address elsewhere.

The specific configuration which has been analyzed is as follows.
Each of the $L=1/2$ vortices is placed at an equal distance $d$
(in units of $\lambda$)
and on opposite sides from the center of the disk, while the straight
domain wall joining them is aligned with the corresponding diameter, for the
reasons given previously. Because of the four-fold symmetry of the 
configuration, it suffices to solve the problem in only one quadrant
of the disk, say for $0\le\phi\le\pi/2$. The lattice discretization
used 20 intervals in the radial direction $0\le u\le u_b$ and
16 intervals in the angular range $0\le\phi\le\pi/2$. Finally, to avoid
as much as possible effects due to the finite boundary---such as the
Bean-Livingston barrier---while still having a sufficiently fine-grained
lattice discretization, the disk radius was set to
\begin{equation}
u_b=5.
\end{equation}

The values for the coupling $\kappa$ which were considered are
\begin{equation}
\kappa\ :\ \ 0.5\ ,\ \frac{1}{\sqrt{2}}\ ,\ 1.0\ ,\ 1.5.
\end{equation}
For each of these values, the steepest descent minimization of the
energy of the system was solved numerically, starting with $d=0$,
as a function of a regularly spaced series of values for the distance $d$,
which measures the position from the center of the disk
of each of the two $1/2$-vortices bound onto the edges of the bound 
$1/2$-domain wall thus streched to the length $2d$.

Fig.2 presents the results of these computations for the value
$\kappa=\kappa_c=1/\sqrt{2}$. The results for the other values
of $\kappa$ show exactly the same behaviour---the only difference being
in the vertical scale for energy values---, with in particular always
a small dip in the energy at the first non zero step in the values
for $d$ whatever the value for $\kappa$. Clearly, the appearance of such
a dip for all the considered values of $\kappa$ does not agree with any
of the expectations discussed above. It is not totally excluded that
this may be due to subtle finite size effects because of the boundary
at $u=u_b$, since $e^{-5}=6.74\cdot 10^{-3}$ is yet not that small. 
Indeed, the total magnetic flux $\Phi[u_b]$
was also monitored during the simulation and does also display
a curious behaviour. When $d=0$, one has a single $L=1$ ANO vortex
at the center of the disk, and $\Phi[u_b]$ then takes a value very
close to unity as it should, namely $\Phi[u_b]=0.989$. However,
when $d$ then increases, the values for $\Phi[u_b]$ first increase
getting even closer to the value of unity, before finally decreasing
as they should when the two $1/2$-vortices move closer to the disk boundary.
On the other hand, it is not clear whether these two results for ${\cal E}$
and $\Phi[u_b]$ could not also be a numerical artifact for small values
of $d$, even though
exactly the same computer code was applied to the situations with $d=0$
and $d\ne 0$. Unfortunately at this stage, it is not possible to
resolve that specific issue, which may only be addressed through a 
full-fledged analysis including a much finer-grained discretization 
which is necessarily very much more computationally intensive.

Nevertheless, note that the numerical solution does display
the expected linear increase in energy with separation, as soon as the
bound $1/2$-domain wall is sufficiently streched, namely when the
two $1/2$-vortices no longer overlap. Indeed, the linear behaviour
sets in around $d\simeq 0.5$ for all considered values of $\kappa$
(recall that $d$ is measured in units
of $\lambda$). But here again, a much finer-grained analysis as a function
of the separation is required.

Hence, it seems fair at this stage to say that the fascinating and important
issue of the possible instability of the ANO vortex as a function of
$\kappa$,
which would then decay into a bound $1/2$-domain wall streched to some
length which should also depend on the value for $\kappa$, 
is still completely open. Even though the
quality of this first numerical analysis is too poor to reach a definite
conclusion as well as a detailed understanding, the evidence which is
available does not exclude the possibility, quite to the contrary.

\section{Conclusions and Outlook}
\label{Sect6}

The fine-grained topological analysis 
which considers the transport of the complex scalar field around all possible
finite contours in spacetime, has thus uncovered a fascinatingly
rich zoology of topological solutions to the Landau-Ginzburg-Higgs equations 
of the abelian U(1) Higgs model. Beyond the well known vortex
states of integer winding number, the new configurations
also include vortices of integer
as well as half-integer winding number bound onto the edges of
domain walls, all such vortex configurations possibly also being surrounded
by annular current flows in the case of bounded spatial domains.
The existence and physical consistency of these new states is related
in a fundamental way to the U(1) local gauge invariance of the model,
leading specifically to the all important fact that
any of these states is to be characterized in terms of some double sheeted 
covering of the plane. The specific details of the physical properties of
these solutions, and of their mathematical construction, have been described
in the Introduction and established in the body of the paper.

Even though the numerical analysis has not yet been able to answer some 
important issues, such as for instance
the identification of the lowest energy configurations
for specific combinations of such states, perhaps the most important
of these cases being the possible instability of the usual 
Abrikosov-Nielsen-Olesen vortex to split into a domain wall with two 
vortices of winding number $L=1/2$ bound onto its edges, the mere existence 
of such solutions raises a host of fascinating potential dynamical
properties. Indeed, it now becomes possible to imagine that under
specific electromagnetic disturbances, the magnetic vortices of systems
described by these equations would not only move around in space,
but could then also be deformed elastically into domain walls with
vortices bound onto their edges, very much like rubber bands which would
be pulled in different directions.
Such properties should have important consequences, for instance, 
with regards to the transport of electromagnetic energy in the vacuum
of systems obeying these equations, even if only in an
effective way.
Quite obviously, many issues of potential great direct physical
interest, both in the static as well as in the dynamical regime, are 
raised by the existence of these new topological states in the abelian 
U(1) Higgs model.

Beyond their physical interest, these solutions raise other questions as well.
{}From the mathematical physics point of view, it would presumably
be very satisfying if the existence of such topological configurations,
as well as of their stability, could be put on a firm 
ma\-the\-ma\-ti\-cal ground,
as has been done for the usual vortex solutions\cite{Taubes,Sigal}. 
One may also consider applying (an adapted form of)
the fine-grained topological analysis to other field theories
known to possess topological states,
and possibly uncover further structure in these states as has been done
here, for example by lifting specific symmetry properties such as rotational 
invariance (indeed, among the states described in this paper, only the
usual vortices possess that symmetry). Extensions to higher spacetime 
dimensions or differential forms of higher degree could also be envisaged,
having in mind the close analogy with D-branes and strings in M-theory,
and the BPS bounds which are saturated for infinitely thin domain walls
and vortices, corresponding then to sets of bound membranes and Aharonov-Bohm 
flux lines.

Closer to particle physics considerations, these new solutions may also
play an in\-te\-res\-ting role in the Landau-Ginzburg model of
dual colour superconductivity as an effective description for the
dual Meissner effect and confinement in QCD. Finally, since vortices
and domain walls are quite generic in grand unified theories with
spontaneous symmetry brea\-king, the possibility that the states
described here, or their cousins, could become involved in some
interesting ways in texture formations\cite{Kibble} in the Universe should 
not be excluded either.

Nonetheless, the field for which the results of this paper offer perhaps
the most im\-me\-dia\-te interest is that of superconductivity.
The issue of the possible instability of the Abrikosov vortex is
especially intriguing. Had we already been able to identify the
equilibrium configuration of the streched bound $1/2$-domain wall
as a function of the Landau-Ginzburg parameter $\kappa$, this could
have been confronted to actual data, say in an Abrikosov lattice or 
in a nanoscopic superconducting disk\footnote{A detailed magnetic field
map of the bound $1/2$-domain wall would also have to be known,
in order to assess the required sensitivity in field values of the
magnetic imaging technique to be used. Spatial resolution however,
is no longer an issue\cite{KUL}}. 
It may even be that the streching of the domain wall could be
amplified to some degree by applying some external magnetic field.
Such questions are certainly fascinating and deserve detailed study. Note 
that the recent results of Ref.\cite{Field} may be related to these issues.
As remarked above, the existence of split vortices and domain walls,
even if not as stable states, should also imply specific dynamical
properties of superconductors submitted to electromagnetic
disturbances, especially in regimes of frequencies whose wave lengths
become comparable to the penetration and coherence lengths, and
to the actual size of vortices and the thickness of domain walls on which 
electromagnetic waves would then pull almost as if rubber bands.

In fact, we were led into the analysis of the issues of this paper
because of our few years old 
project to develop a quantum detector for polarized
particles based on superconducting loops\cite{UCL}. Indeed, when such a loop
is polarized in an external magnetic field close enough to the switching
between two successive winding number states so that some local disturbance
in the magnetic field would cause the loop to switch from one state to the
other and back, this non linear response of the loop by a significant
fraction of a full quantum of flux may be picked up by an inductive circuit 
coupled to a SQUID.
The same principle is at work in other devices using superconducting
loops, albeit of larger dimensions,
aimed towards the development of superconducting\cite{Likharev} (or) 
quantum\cite{Friedman} computers. However, in order to optimize the 
geometry of the loop
with the particle detector application in mind, it becomes necessary
to have a detailed understanding of the dynamics in a relativistic
regime of the switching mechanism, which seems not be known.
Only energy based considerations in the static regime are available
for superconducting mesoscopic disks and annuli, mostly motivated
by the results of Ref.\cite{Geim}. The only exception may be the analysis
of Ref.\cite{Berger} which provides a detailed picture for how an integer vortex
may move in and out a superconducting disk or loop as a function of the
applied field. Now that $1/2$-vortices bound onto the edge of a $1/2$-domain
wall have been uncovered from the same Landau-Ginzburg equations,
the whole issue has to reconsidered again, to see whether there are
specific loop geometries for which the sensitivity of the dynamical
response to disturbances could be optimized, for example by having the
$L=1/2$ state have a lower energy than the $L=1$ one when being
degenerate with the $L=0$ state, so that magnetic flux would move in and out
of such devices through a string of smaller bits than when only
integer vortices are involved.

Yet once again, the abelian U(1) Higgs model has proved to have still
unknown precious tools hidden at the bottom of its treasure chest,
to provide us, as if the Robinson Crusoes of the Universe stranded on our
little ``speckle of dust", with the means to explore ever further the
physical world at so many different scales.

\vspace{20pt}

\noindent{\bf Acknowledgements.}
It is a pleasure to acknowledge Olivier van der Aa and Geoffrey Stenuit 
for constructive discussions concerning numerical simulations.
The author also wishes to thank Alfred Goldhaber 
and especially Martin Ro\v{c}ek, Gary Shiu and Stefan Vandoren for
their interest and for discussions, as well as
Peter van Nieuwenhuizen and the members of the
C.N. Yang Institute for Theoretical Physics at the State University of
New York at Stony Brook for their hospitality and a stimulating
working environment.

\clearpage

\noindent{\bf Figure Captions}

\vspace{10pt}

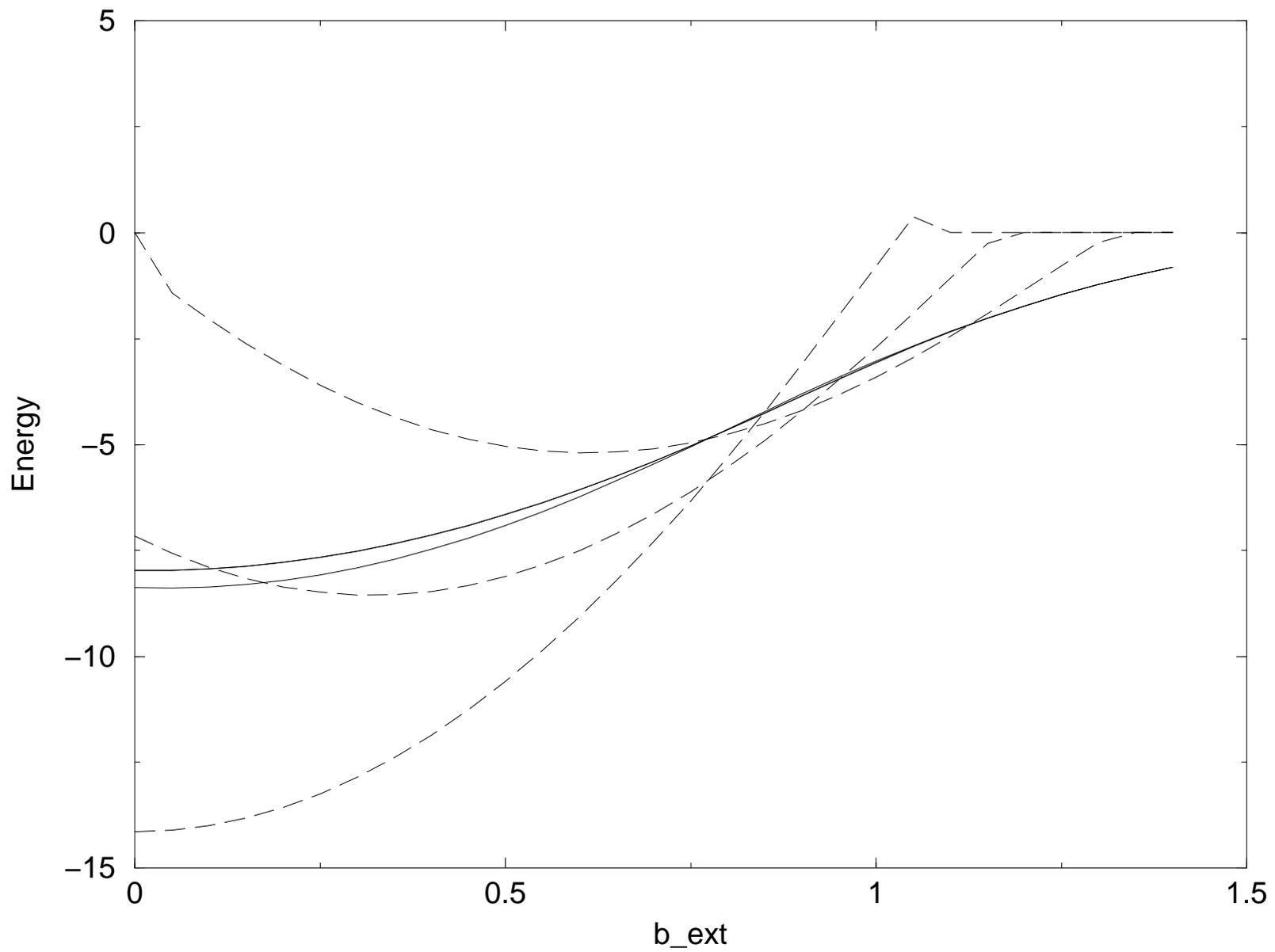
\noindent Figure 1: The free energy ${\cal E}$
of the disk as a function of the applied
external magnetic field, for $u_b=3$ and $\kappa=1$. The vortex configurations
are those discussed in section \ref{Subsect4.3}, with the vortices
at the center of the disk and having winding numbers $L=0,1,2$ and $L=1/2,3/2$,
including in the latter two cases a single domain wall extending up to the disk 
boundary. The three dashed lines correspond, from bottom to top, to the states
with $L=0,1,2$ in the same order, while the two continuous lines correspond,
in the same order again, to the states with $L=1/2,3/2$. For further
details, see section \ref{Subsect4.3}.

\vspace{20pt}

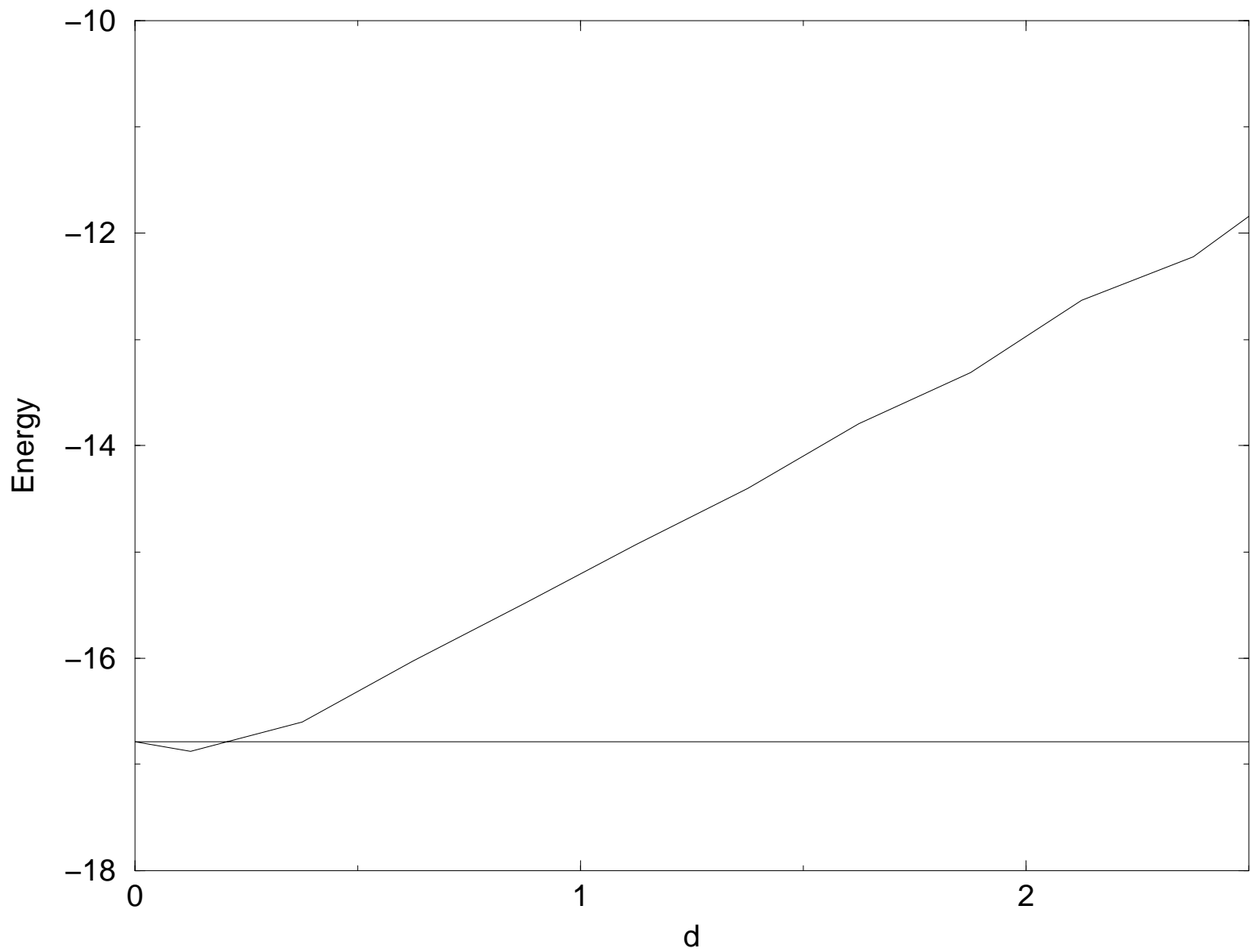
\noindent Figure 2: The energy ${\cal E}$
of the bound $1/2$-domain wall in a disk
with $u_b=5$ and $\kappa=1/\sqrt{2}$ as a function of the distance $d$ to 
the disk center of each of the $1/2$-vortices bound onto the edges of the
domain wall. For further details, see section \ref{Subsect5.2}. The
horizontal line is a guide to the eye, and corresponds to the
value for $d=0$.

\end{document}